\documentstyle{jfm}
\input psfig.sty

\title{Rotations and cessations of the large-scale circulation in turbulent Rayleigh-B{\'e}nard convection}
\author{Eric Brown and Guenter Ahlers}
\affiliation{Department of Physics and iQCD, University of California, Santa Barbara, CA 93106}

\begin{document}

\maketitle
\date{\today}

\begin{abstract}
We present a broad range of measurements of the angular orientation $\theta_0(t)$ of the large-scale circulation (LSC) of turbulent Rayleigh-B{\'e}nard convection  as a function of time.  We used two cylindrical samples of different overall sizes, but each  with its diameter nearly equal to its height.  The fluid was water with a Prandtl number of 4.38. The time series $\theta_0(t)$ consisted of  meanderings similar to a diffusive process, but in addition contained large and irregular spontaneous reorientation events through angles $\Delta \theta$. We found that reorientations can occur by two distinct mechanisms. One consists of a rotation of the circulation plane without any major reduction of the circulation strength. The other involves a cessation of the circulation, followed by a restart in a randomly chosen new direction. Rotations occurred an order of magnitude more frequently than cessations. Rotations occurred with a monotonically decreasing  probability distribution $p(\Delta \theta)$, {\it i.e.} there was no dominant value of $\Delta \theta$ and small $\Delta \theta$ were more common than large ones.  For cessations $p(\Delta\theta)$ was uniform, suggesting that information of $\theta_0(t)$ before the cessation is lost.  Both rotations and cessations have Poissonian statistics in time, and can occur at any $\theta_0$.  The average azimuthal rotation rate $|\dot\theta|$ increased as the circulation strength of the LSC decreased. Tilting the sample relative to gravity significantly reduced the frequency of occurrence of both rotations and cessations.

\end{abstract}

\section{Introduction}

The problem of Rayleigh-B{\'e}nard convection (RBC) consists of a fluid sample heated from below [for example, see \cite{Si94, Ka01, AGL02}].  The heat drives a convective flow and is thus transported out the top of the sample.  In our case the sample is a cylindrical container filled with water.  This system is defined by three parameters:  the Rayleigh number $R \equiv \alpha g \Delta T L^3/\kappa \nu$ ($\alpha$ is the isobaric thermal expansion coefficient, $g$ the acceleration of gravity, $\Delta T$ the applied temperature difference, $L$ the height of the sample, $\kappa$ the thermal diffusivity, and $\nu$ the kinematic viscosity), the Prandtl number $\sigma \equiv \nu / \kappa$, and the aspect ratio $\Gamma \equiv D/L$ ($D$ is the diameter of the sample).  Convection happens as a result of the emission of volumes of hot fluid known as ``plumes" from  a bottom thermal boundary layer that rise due to a buoyant force, while cold plumes emitted from a top boundary layer sink.  In the turbulent regime of $\Gamma=1$ samples that we study, these plumes drive a large-scale circulation (LSC) [\cite{KH81, SWL89, CGHKLTWZZ89, CCL96, QT01a, FA04, SXT05,TMMS05}], also known as the ``mean wind",  which is oriented nearly vertically with up-flow and down-flow on opposite sides of the sample.

  The LSC configuration does not have the rotational invariance of the cylindrical sample, but the cylindrical symmetry implies that any azimuthal orientation $\theta_0$ of the LSC is an equally valid state for the system.  In this paper we present extensive measurements of spontaneous angular changes $\Delta \theta$ of $\theta_0$, i.e. of reorientations of the LSC. Such changes have been observed previously [\cite{Ke66, We67, CDBS75, GWR84, HYK91, CCS97, NSSD01, FO02, SBN02,BNA05a, SXX05, XZX06}].   In one case, a {\it rotation} of the entire structure through an angle of $\Delta\theta \simeq 1/2$ revolution without a significant change in flow speed was clearly observed in an experiment using mercury as the fluid [\cite{CCS97}].  Another conceivable mechanism is {\it cessation}, in which the LSC flow-speed vanishes, and then the flow restarts in a different direction.  Cessation was observed in numerical simulations [\cite{HYK91,FO02}],  a dynamical-systems model [\cite{AGL05}], and a stochastic model [\cite{Be05}]. All of these cases are two-dimensional models where only cessations with $\Delta\theta = 1/2$ revolution are possible, although in principle the models could be extended to three dimensions.  Cessation also occurs in convection loops (a thin circular vertically oriented loop filled with fluid heated in the lower and cooled in the upper  half) where because of the two-dimensional nature again only cessations with $\Delta\theta = 1/2$ revolution are possible [\cite{Ke66, We67,CDBS75,GWR84}]. 
  
  The experimental work by \cite{NSSD01} and subsequent analysis by \cite{SBN02} yielded statistics relating to reversals of the LSC, but could not determine $\Delta \theta$ and was unable to distinguish between the rotation and cessation mechanisms. Cessations were first documented in a laboratory sample of turbulent RBC by \cite{BNA05a}, and the present  paper presents more extensive analysis and additional data from that project.  Recent experiments in $\Gamma = 0.5$ samples [\cite{SXX05}] measured the orientation of the LSC, and contemporary experiments in $\Gamma = 1$ samples [\cite{XZX06}] have produced results complementary to our own regarding the azimuthal dynamics of the LSC.  With these experiments, the azimuthal dynamics of the LSC are beginning to be understood.  Spontaneous changes of the orientation of the LSC are not only interesting from the standpoint of fundamental physics, but are important in many geophysical applications.  For instance, reversals occur in natural convection of the Earth's atmosphere [\cite{DDSC00}].  Convection dynamics in the outer core of the Earth are responsible for changes in the orientation of Earth's magnetic field [\cite{GCHR99}].

The goal of the present work is to better understand the reorientations of the LSC, by both the rotation and cessation mechanisms.  We first explain the experiment and how we determine $\theta_0$ in Sect.~\ref{sec:experiment}.  In Sect.~\ref{sec:reorientations} we illustrate the existence and nature of both rotations and cessations. In Sect.~\ref{sec:reorientationstat} we present statistics relating to reorientations of the LSC, showing that successive rotations are independent of each other, and that they result in a wide range of $\Delta\theta$, in which strict reversals ($\Delta\theta \simeq 1/2$ rev.) are not especially common.  In Sect.~\ref{sec:cessation} we present statistics relating to cessations. These are found to be an order of magnitude more rare than rotations. We show that after the LSC stops, it restarts with a random new orientation.  In Sect.~\ref{sec:comparison} we compare our results with those of Sreenivasan and coworkers. There we show that we can reproduce the statistics that they derived from their data only when we count an event  each time when the orientation crosses a fixed angle. We refer to such events as ``crossings". Crossings include events caused by small-amplitude, high-frequency ``jitter" near the crossing angle, and considering them resolves apparent inconsistencies between the two experiments.  In Sect.~\ref{sec:dthetadt} we present statistics of the azimuthal rotation rate over the long term. There we  show that the angular distance travelled by $\theta_0$ scales as in a diffusive process, and that the absolute value of the rotation rate  $|\dot\theta|$ increases when the LSC amplitude decreases.   In Sect.~\ref{sec:tilt} we present results from tilting the sample relative to gravity.  These data complement results already reported by \cite{ABN05}. Since naturally occurring convection systems are generally not cylindrically symmetric, it is important to study how rotations and cessations behave in less symmetric systems.   As observed by others (see, for instance, \cite{SXT05} and references therein), we find that the tilt pushes the LSC into a preferred orientation along the direction of the tilt. In addition, we determined that both rotations and cessations are strongly suppressed by the tilt.  Finally in Sect.~\ref{sec:Xi}, we comment on the contemporary experiments by \cite{XZX06}, who reported on the azimuthal dynamics of the LSC, and compare the results to ours. A brief summary is given in Sect.~\ref{sec:summary}. 

Because the theoretical models [\cite{AGL05, Be05}] only predict two-dimensional reversals, and not the three-dimensional rotations and cessations that we observe, they cannot yet be compared in detail to the experimental data.

\section{The apparatus and experimental method}
\label{sec:experiment}

The experiments were done with two cylindrical samples with aspect ratio $\Gamma  \simeq 1$ that are the medium and large sample described in detail elsewhere [\cite{BNFA05}].  Both had circular copper top and bottom plates with a plexiglas side wall that fit into a groove in each plate.   There were no internal flanges, seams, sensors, or other structures that could interfere with the fluid flow.  The medium sample had  $D=24.81$ cm and $L=24.76$ cm, and the large sample had $D = 49.67$ cm and $L = 50.61$ cm.
Each sample was filled with water and the average temperature between the bottom and top plates was kept at $40.0^{\circ}$ C where $\sigma = 4.38$. The two samples of different heights allowed us to cover a larger range of $R$ at the same $\sigma$ and $\Gamma$, so the overall range studied was $3\times10^8 \stackrel{<}{_\sim} R \stackrel {<}{_\sim} 10^{11}$.  Three rows of eight blind holes each, equally spaced azimuthally and lined up vertically with each other at heights $3L/4$, $L/2$, and $L/4$,  were drilled from the outside into the side walls of both samples. Thermistors were placed into them so as to be within $d = 0.07$ cm of the fluid surface.  Earlier experiments were done with only the middle row of eight thermistors at height $L/2$, and presentations of data in this paper that mention only one row of measurements refer to the middle row, which was sampled in both the early and later experiments.  Since the LSC carried warm (cold) fluid from the bottom (top) plate up (down) the side wall, these thermistors detected the location of the upflow (downflow)  of the LSC by indicating a relatively high (low) temperature.  No parts of the thermistors extended into the sample where they might have  perturbed the flow structure of the fluid.  The lead wires for these thermistors were wrapped around the insulating layers just outside the side wall to prevent the introduction of heat currents into the sides of the samples from these leads.  The thermistors have a resolution of $10^{-3}$ degrees C. Both samples were carefully levelled to better than 0.001 rad, except for the experiments in which we deliberately tilted the samples.

We presume that the side-wall thermistors measured the temperature of the thermal plumes and the accompanying LSC, just outside of the viscous boundary layer at the side wall.  Here we address the issue of whether one can measure these temperatures through the boundary layer.  The response time of the thermistors for thermal diffusion through the side wall is $d^2/\kappa_{sw} = {\cal O}(1)$ s, where $\kappa_{sw}$ is the thermal diffusivity of the plexiglas wall.  The response time for heat flow through the viscous boundary layer -- assumed to have a width $\lambda = 0.464L\times R_e^{-1/2}$ [\cite{GL02} with fit parameter from \cite{BFA06}] --  for thermal diffusion is expected to be $\lambda^2/\kappa \approx 0.215L^2/(\kappa R_e)$. For the large sample this time ranged from about 140s for $\Delta T \simeq 1^\circ$C ($R_e \simeq 2500$) to about 32s for  $\Delta T \simeq 20^\circ$C ($R_e \simeq 11000$). 
However, this calculation greatly overestimates the real response time, since the data in this paper show that we could observe temperature changes that occurred over time scales as short as several seconds.  Consistent with direct measurements of fluctuations in the viscous boundary layer [\cite{QX98}], the relatively fast thermal response time suggests that there is turbulent mixing in this boundary layer that enhances the heat transport to the side-wall thermistors.

\begin{figure}
\centerline{\psfig{file=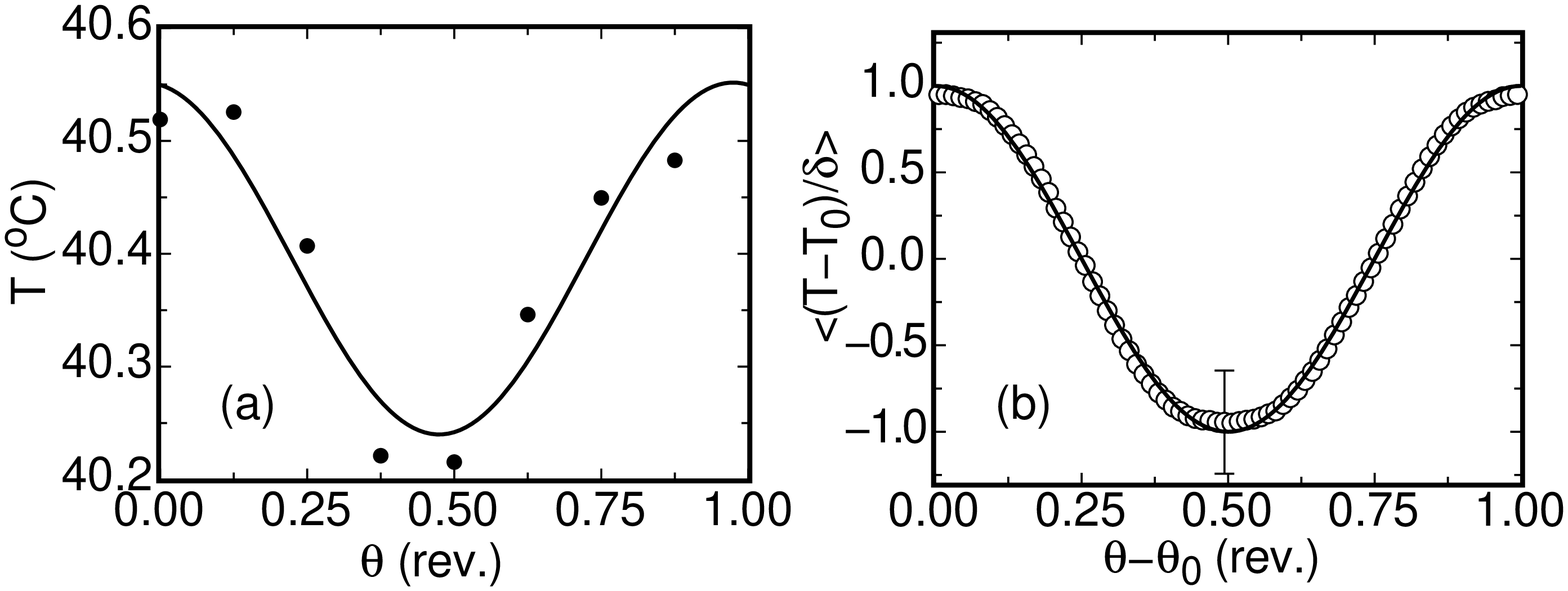,width=5.1in}}
%\vskip 0.2in
\caption{(a):  An example from the large sample of the temperatures at the horizontal mid-plane of the side wall as a function of  the azimuthal angle $\theta$ for $R = 9.6 \times 10^{10}$. Solid line: a fit of $T_i = T_0 + \delta\cos(i \pi/4 - \theta_0), i = 0, \ldots,7$ to the data. The fit yields the orientation $\theta_0$  and an amplitude $\delta$ that reflects the LSC. (b):  The averaged normalized side-wall-temperature-profile $(T_i-T_0)/\delta$, sorted into bins with each bin covering a small range of $\theta-\theta_0$.  Solid line: a cosine function.  Vertical bar:  typical sample standard deviation for each bin.}
\label{fig:temp_theta}
\end{figure}

We made measurements with a sampling period $\delta t$ as short as 2.5 seconds, and fit the empirical function 
\begin{equation}
T_i = T_0 + \delta\cos(i \pi/4 - \theta_0),\  i = 0, \ldots,7\ ,
\label{eq:T_i}
\end{equation}
 separately at each time step, to the eight middle-row side-wall thermistor-temperature readings. An example of such a fit is shown in Fig.~\ref{fig:temp_theta}a.  Deviations from a smooth profile are presumed to be due to the turbulent nature of the system; for instance a hot plume passing by a thermistor will cause a higher than average temperature reading at that particular angular location. The fit parameter $\delta$ is a measure of the amplitude of the LSC and $\theta_0$ is the azimuthal orientation of the plane of the LSC circulation.   As defined here, the orientation $\theta_0$ is on the side of the sample where the LSC is warm and up-flowing and is measured relative to the location of thermometer zero, which was located on the east side of the sample.  Typically the uncertainties for a single measurement were about 13\% for $\delta$ and 0.02 rev. for $\theta_0$.   Fitting to the cosine function does not yield a unique $\theta_0$ because it is $2\pi$ periodic.  The final value of $\theta_0$ was chosen as the one closest to $\theta_0$ of the previous timestep, thus allowing us to observe rotations of the LSC through larger angles, than if we had reduced the range to $0 < \theta_0 < 2\pi$.  We calculated orientations $\theta_t$ and $\theta_b$ and amplitudes $\delta_t$ and $\delta_b$ for the top and bottom rows by the same method as for the middle row.  Throughout the paper, when we show data based on one row of thermistors it is for the middle row, while the top and bottom row measurements are only used when data is shown for all three levels.

To test the validity of the sinusoidal fitting function, Fig.~\ref{fig:temp_theta}b shows the side-wall temperature-profile normalized by the fit values.  Each point is an average of the normalized termperature $(T_i-T_0)/\delta$ in a bin with a small range of $\theta - \theta_0$.  The standard deviation of the normalized temperature for each bin is about 0.30 and nearly independent of $\theta-\theta_0$.  It is shown as a vertical bar in the plot to indicate the typical size of temperature fluctuations.  The data is in good agreement with a cosine function (solid line) without any additional fitting, indicating that Eq.~\ref{eq:T_i} is a good function to represent the average temperature profile around the side wall relative to $\theta_0$.

The side-wall thermistors are also used to obtain the plume turnover time.  Autocorrelations of a single side-wall thermistor-temperature yielded peaks at times $0, {\cal T}, 2{\cal T}$, etc., while cross-correlation functions of mid-plane thermistors on opposite sides of the sample were negative and yielded peaks at times ${\cal T}/2,  3{\cal T}/2$, etc.  The peaks in the correlation functions indicate a periodicity in temperature fluctuations, i.e. plumes, circulating in the sample, hence we call ${\cal T}$ the plume turnover time.  The technical details of this measurement will be reported by \cite{BFA06}.

\begin{figure}
\centerline{\psfig{file=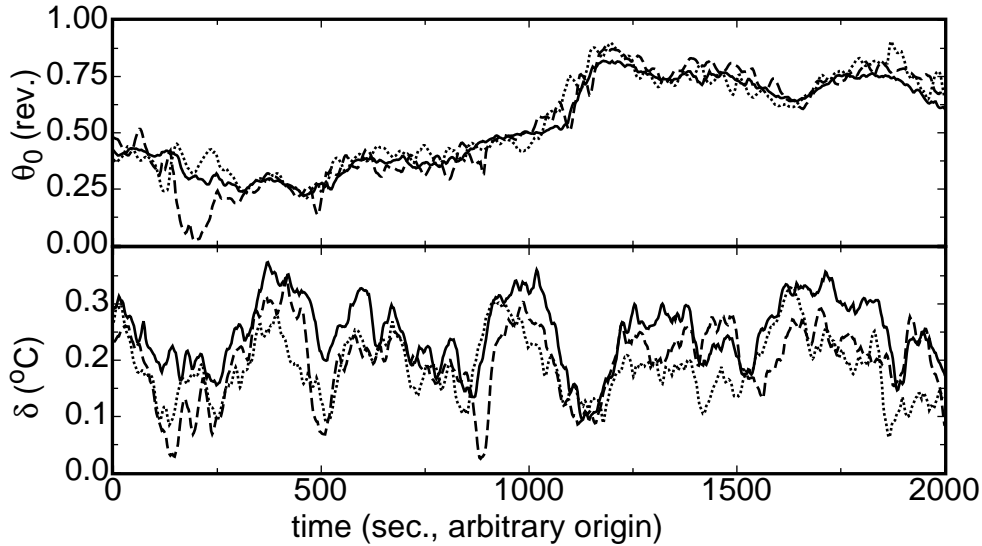,width=5.1in}}
%\vskip 0.2in
\caption{A time series of the orientation $\theta_0$ and amplitude $\delta$ showing rotations of the LSC.  Data is from the medium sample at $R=1.1\times 10^{10}$.  Solid line:  middle-row thermistors.  Dashed line:  top-row thermistors.  Dotted line:  bottom-row thermistors.  }
\label{fig:reorientation_examples}
\end{figure}

\section{The nature of rotations and cessations}
\label{sec:reorientations}

We previously published time series of $\theta_0$ and $\delta$ for the samples with 8 side-wall thermistors [\cite{BNA05a}].  Figure \ref{fig:reorientation_examples} shows a half-hour time series of $\theta_0$ and $\delta$ from the medium sample with 24 side-wall thermistors at $R=1.1\times 10^{10}$.  The orientation and amplitudes are shown separately for the three rows of thermistors.  The data contain a series of erratic rotations of the orientation of the LSC.  Through these rotations, a reversal of the LSC direction ($\Delta\theta \simeq 1/2$ rev.) is made over several hundred seconds, roughly during the time interval from 500 to 1200s.  This is a slow reversal relative to the plume turnover time ${\cal T}$, which is 49 seconds in this case.   This type of reversal by rotation has been reported before, using temperature measurements at various azimuthal locations in the bottom plate of a convection cell to determine the LSC orientation [\cite{CCS97}].  It is important to note that throughout these rotations, the temperature amplitude $\delta$ remains non-zero.  This implies that the LSC was circulating over this entire period, so these are rotations and not cessations.  The values for the three rows are seen to agree fairly well, which agrees with our assumptions about the vertical alignment of the LSC, but the top- and bottom-row temperatures generally have more variability than the middle-row temperatures, and the amplitudes $\delta_t$ and $\delta_b$ drop on occasion without significantly affecting the middle-row signal.

\begin{figure}
\centerline{\psfig{file=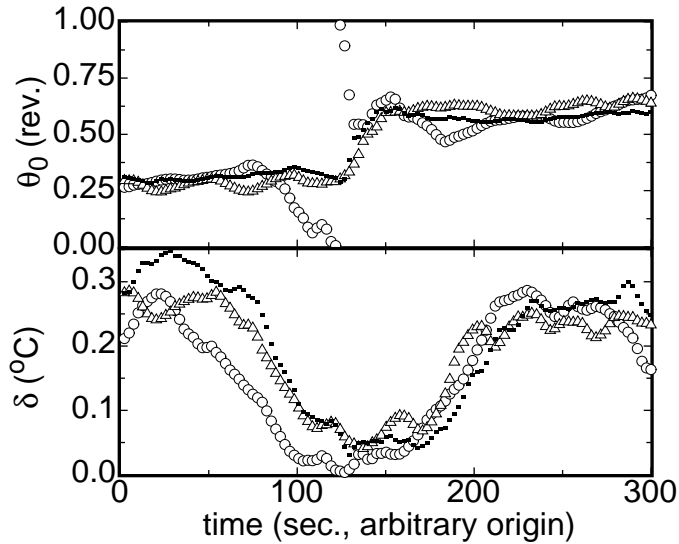,width=3.5in}}
%\vskip 0.2in
\caption{A time series of the orientation $\theta_0$ and of the amplitude $\delta$ showing a cessation of the LSC, in which the amplitude drops to zero.  Data is from the medium sample at $R=1.1\times 10^{10}$.  Solid squares:  middle-row thermistors.  Open triangles:  top-row thermistors.  Open circles:  bottom-row thermistors.}
\label{fig:cessation_examples}
\end{figure}

Figure \ref{fig:cessation_examples} shows another time series of $\theta_0$ and $\delta$ for the three rows of thermistors.  Here $\delta$ decreased essentially to zero and then increased back up close to its average value.  We interpret an amplitude drop to also indicate a velocity drop, since the temperature distribution -- represented by $\delta$ --  drives the LSC by buoyancy, and experiments have found a correlation between temperature and velocity [\cite{NSSD01, QSTX04}].  This means that the LSC gradually slowed to a stop, reversed direction, and gradually sped up again in another direction without significant rotation of the plane of circulation.  This is clearly a cessation, but note that it is not strictly a reversal because $\Delta\theta < 1/2$ rev.

\section{Reorientation statistics}
\label{sec:reorientationstat}

In this section we examine the statistics of reorientations, regardless of whether they occur by rotation or cessation. However, these results reflect primarily the properties of rotations because, as we shall show below in Figs.~\ref{fig:reor_rate} and \ref{fig:cess_rate}, cessations are an order of magnitude more rare than rotations. We encountered a reorientation event roughly once per hour (see Fig.~\ref{fig:reor_rate} below). The statistics of cessations will be discussed separately in Sect.~\ref{sec:cessation}.

\subsection{Definition of reorientations} 

Because rotations do not necessarily have clear starting and ending points, and may have a wide range of sizes and speeds, they are difficult to define for the purpose of data processing and statistical analysis.  Instead, we start by defining a {\it reorientation} as an event with a sufficiently large and quick change in the orientation of the LSC.   More specifically, we required reorientations to satisfy two criteria, using only data from the middle row of thermistors since that was available from all of the experiments. These criteria are the same as those used by \cite{BNA05a}.  First, the magnitude of the net angular change in orientation $| \Delta\theta |$ over a set of successive data points for $\theta_0$ had to be greater than a chosen parameter $\Delta\theta_{min}$. Second,  the magnitude of the net average azimuthal rotation rate $|\dot\theta| \equiv |\Delta \theta / \Delta t|$ over that set had to be greater than a chosen parameter $\dot\theta_{min}$.   Here $\Delta t$ is the duration of the reorientation. Usually multiple overlapping sets satisfied these requirements, so in those cases the set with the maximum local reorientation quality factor $Q_n = |\Delta\theta| / (\Delta t)^n$ was chosen as the reorientation.  For $0 < n < 1$, $Q_n$ represents a compromise between choosing the maximum angular change ($Q_0$) or the maximum rotation rate ($Q_1$).  Any adjacent points to the chosen set were also included if the instantaneous rotation rate $\dot\theta_{0} = \delta\theta_0 / \delta t$ [$\delta\theta_0 = \theta_0(t+\delta t) - \theta_0(t)$] for the adjacent point was greater than $\dot\theta_{min}$ and of the same sign as for the reorientation.  For the results presented in this paper, we used the parameters $\Delta\theta_{min} = 0.125$ rev.,  $\dot\theta_{min} = (0.1 \mbox{ rev. }) /  {\cal T}$, and $n = 0.25$; but since all three reorientation definition parameters are arbitrary, we did the analysis over the ranges  $1/64 \mbox{ rev. } \le \Delta\theta_{min} \le 0.25 \mbox{ rev. }$,  $0.0125 \mbox{ rev. } \le \dot\theta_{min} {\cal T} \le 0.2 \mbox{ rev. } $, and $0 \le n \le 1$ to confirm that the physical results were not sensitive to the choice of reorientation definition parameters. For $\Delta\theta_{min}$ and $\dot\theta_{min}$, the smallest values used were on the order of the turbulent fluctuations in the data, so nearly every data point would be counted as a reorientation, and at the largest parameter values used too few events were counted to yield useful statistics.  The qualitative conclusions of the analysis did not change with the different parameter values we tried. Some of the quantitative values found from fitting statistical distributions did change.  Any variation of these fit values with the chosen parameters will be mentioned where it is appropriate.

\begin{figure}
\centerline{\psfig{file=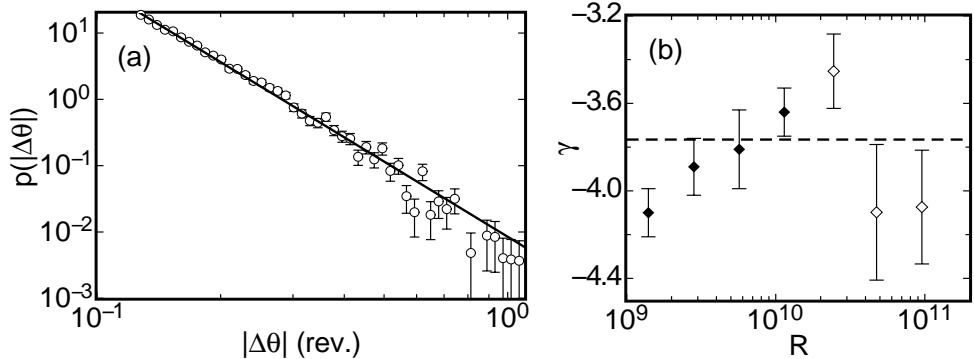,width=5.1in}}
\caption{(a):  Probability distribution $p(|\Delta\theta|)$ of the angular change $\Delta\theta$ for reorientations.  Open circles: experimental data. Solid line: power-law fit to the data.  (b): The exponent $\gamma$ from the power-law fits of $p(|\Delta\theta|)$ as a function of $R$ for the medium sample  (solid diamonds) and large sample (open diamonds). Dashed line:  the value of $\gamma$ from the fit to all data.}
\label{fig:prob_delta_theta}
\end{figure}

\subsection{Results}

Probably the most important feature of reorientations is the angular change $\Delta\theta$.  The probability distribution $p(|\Delta\theta|)$ is shown in Fig.~\ref{fig:prob_delta_theta}a for all data, regardless of $R$.  Here the data were sorted into bins that were evenly spaced on a logarithmic scale.  The error bars represent the probable error of the mean,  with the relative error of the mean taken to be the inverse square root of the number of reorientations in a bin.  Fitting a power law $p(|\Delta\theta|) \propto \left(|\Delta\theta|\right)^{\gamma}$ to the data yielded $\gamma = -3.77\pm0.04$.  The fit was done by the maximum-likelihood method [see, for instance, \cite{BR92}] to avoid errors associated with the binning of the data.  The same analysis was done also for reorientations seperately at various $R$.  These probability distributions were again fitted by power laws, with the resulting $\gamma$ values shown for each $R$ in Fig.~\ref{fig:prob_delta_theta}b. The exponent of the distribution was, within our resolution, independent of $R$.  The power-law exponents found could vary by up to a factor of 2 for extreme values of the reorientation definition parameters.  However, for all  parameter values tried, the distribution was consistent with a power law with a negative exponent that did not vary significantly with $R$, and there were never any peaks in the distribution.  The major conclusions are that there is a monotonically decreasing distribution of $|\Delta \theta|$, so that smaller reorientations are much more common than larger ones, and that there is no characteristic reorientation size.  This analysis shows that the strict reversal is not especially common among reorientation events.  The probability distribution in Fig.~\ref{fig:prob_delta_theta} implies that only about 1\% of the reorientations we counted have $|\Delta\theta|=1/2 \pm 0.05$ rev.  This result is significant because an interpretation of previous experimental work had suggested that the events found were all reversals [\cite{NSSD01}], and two-dimensional theoretical models can predict only reversals [\cite{AGL05, Be05}].

\begin{figure}
\centerline{\psfig{file=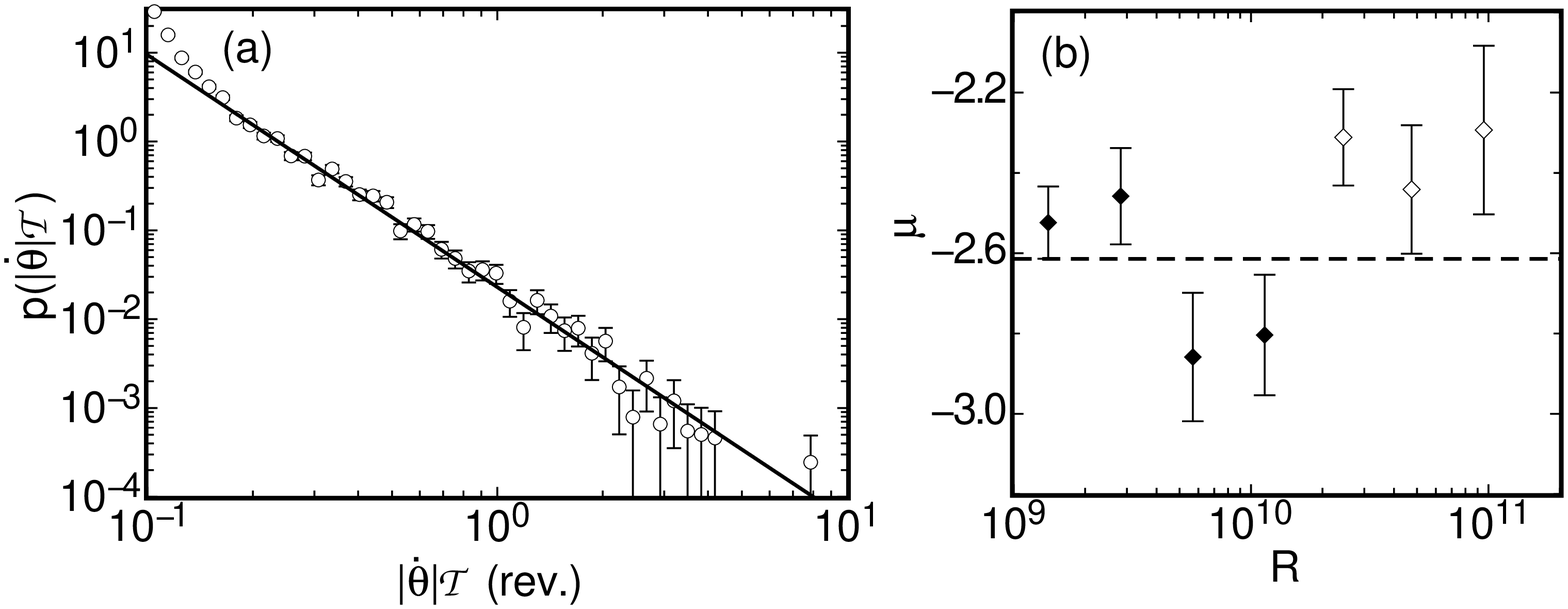,width=5.1in}}
\caption{(a): The probability distribution $p(|\dot\theta| {\cal T})$ of the azimuthal rotation rate for reorientations at all $R$. Open circles: experimental data. Solid line: power-law fit to the data. (b): The exponent $\mu$ from the power-law fit of $p(|\dot\theta|{\cal T})$ as a function of  $R$ for the medium sample  (solid diamonds) and large sample (open diamonds).  Dashed line:  the value of $\mu$ from the fit to all data.}
\label{fig:reor_prob_dthetadt}
\end{figure}

The average azimuthal rotation rate $|\dot\theta|$ for reorientations was studied by a method analogous to that used for $|\Delta \theta|$.  All of the reorientations were sorted into bins according to $|\dot\theta|\cal T$, to make the rotation rate dimensionless so data at different $R$ could be compared. The probability distribution $p(|\dot\theta| {\cal T})$ is plotted in Fig.~\ref{fig:reor_prob_dthetadt}a, with error bars equal to the probable error of the mean.  Fitting a power law to the data in the range $|\dot\theta| {\cal T} \ge 0.16$ rev. yielded $p(|\dot\theta| {\cal T}) \propto (|\dot\theta| {\cal T})^{\mu}$, where $\mu = -2.61\pm0.04$.  The same analysis was also done for individual values of $R$, and $\mu$ is shown for each in Fig.~\ref{fig:reor_prob_dthetadt}b.  The exponent $\mu$ can vary by about a factor of 2 with extreme values for the reorientation definition parameters, but the qualitative results are unchanged.  The probability distribution $p(|\dot\theta| {\cal T})$ is described well by a power law with a large negative exponent that is within our resolution independent of $R$, showing that slower reorientations are much more common than faster ones, and that there is no characteristic rotation rate for reorientations. We note that the normalization by ${\cal T}$ is not independent of $R$. In fact, ${\cal T} \propto R^{-0.5}$ [\cite{BFA06}] over most of our range of $R$.

The probability distribution of the duration $\Delta t$ of reorientations, which is not shown here,  is sharply peaked, and the peak location coincides with $\Delta\theta_{min} / \dot\theta_{min}$.  This is approximately true for all of the values of the reorientation parameters we studied.   Since this characteristic duration always depended on these artificial parameters, there seems to be no characteristic physical time scale for the duration of reorientations that we could measure.  Inspection of the data indicates that the duration of reorientations can be as short as $0.1{\cal T}$ (these appear to be cessations) and the longest durations seem to be about equal to the artificial value $(1 \mbox{ rev.})/\dot\theta_{min}$.

Since the method for defining and counting reorientations is non-traditional, the analysis program was run on a simulated Brownian diffusive process for comparison.  The orientation $\theta_0$ was allowed to travel in one dimension in discrete steps at time intervals equal to the sampling interval $\delta t$ of experiments.  The orientation change $\delta\theta_0$ for each time step was randomly generated with the distribution of Gaussian white noise.  The simulated noise was made to have the same root-mean-square step size as the diffusive region of the real data for $R=1.1\times 10^{10}$ (given by $\varphi\sqrt{\delta t}$ as defined in Sect.~\ref{sec:dthetadt}). For this simulated data there are about half as many reorientations as in the real data.  While $p(|\Delta\theta|)$ looks similar in shape to that for the real data, it falls below the real data and an exponential distribution fits better than a power law to $p(|\Delta\theta|)$.  This suggests that the  LSC undergoes significantly more large reorientation events than a diffusive process.

\begin{figure}
\centerline{\psfig{file=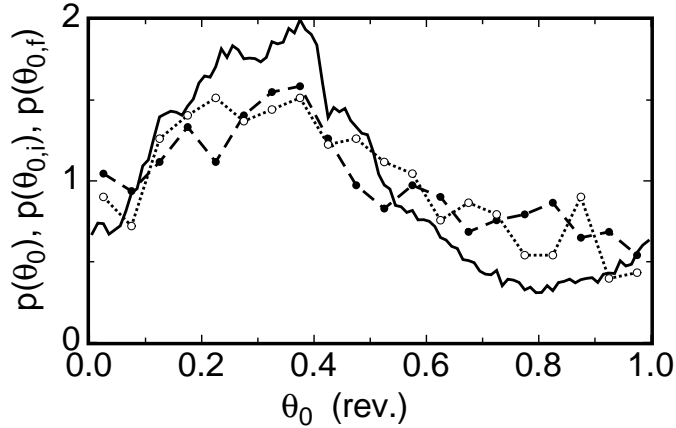,width=3.5in}}
\caption{Solid line: Probability distribution of the orientation of the mean wind $p(\theta_0)$ as a function of $\theta_0$.  Open circles and dotted line: Probability distribution of the starting orientation $p(\theta_{0,i})$ for reorientations.  Solid circles and dashed line:  Probability distribution of the ending orientation $p(\theta_{0,f})$ for reorientations. Data is for $R=1.1\times 10^{10}$ in the medium sample.}
\label{fig:prob_theta}
\end{figure}

One notable point is that there is a preferred orientation $\theta_m$ of the LSC, which is apparent in the probability distribution of the mean wind orientation $p(\theta_0)$ shown in Fig.~\ref{fig:prob_theta} for $R=1.1\times 10^{10}$ (solid line).  In this case, we reduced the orientation to the range $0 < \theta_0 < 2\pi$.  For other $R$, $p(\theta_0)$ is generally found to have a single broad peak at a $\theta_m$ that varies with $R$, but that is reproducible when experiments are done in the same apparatus at the same $R$.  This distribution would ideally be uniform in an azimuthally symmetric system, but minor deviations from perfect rotational symetry such as a slight tilt of the sample relative to gravity, a slightly elliptical cross section of the side wall, or a coupling of the Earth's Coriolis force to the LSC [\cite{BA06}] could cause a deviation from the uniform distribution.  Also shown in the figure are the probability distributions of the starting orientations $p(\theta_{0,i})$ (dotted line) and ending orientations $p(\theta_{0,f})$ (dashed line) of reorientations from the same data set.  Both of these distributions fall reasonably close to $p(\theta_0)$.  This shows that reorientations can occur at any orientation $\theta_0$ of the mean wind.   This conclusion differs from that of \cite{NSSD01}, whose interpretation of their data implied a bimodal distribution of the mean wind orientation, with the orientation switching between two opposite orientations.  This difference will be discussed further in Sect.~\ref{sec:comparison}.

\begin{figure}
\centerline{\psfig{file=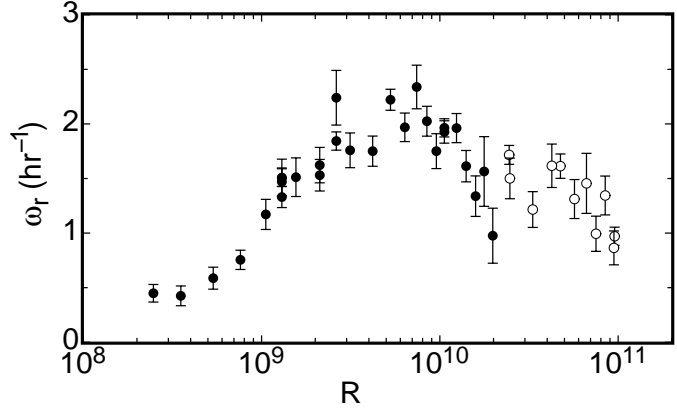,width=3.5in}}
\caption{The average rate of occurrence $\omega_r$ for reorientation as a function of the Rayleigh number $R$.  Solid circles: medium sample.  Open circles: large sample.}
\label{fig:reor_rate}
\end{figure}

The average rate of occurrence $\omega_r$ of reorientations is shown as a function of $R$ in Fig.~\ref{fig:reor_rate}, with error bars equal to the probable error of the mean. This rate did not vary much with $R$ for most of the range studied, but note that since $\dot\theta_{min} = (0.1 \mbox{ rev. })/{\cal T}(R)$ , the minimum requirement is more stringent at larger $R$, so a definition for reorientations with $\dot\theta_{min}$ independent of $R$ would result in the frequency of events increasing with $R$.  The rate $\omega_r$ depended strongly on the reorientation definition parameters: it decreased by about 2.1\% with each increase of 1\% in $\Delta\theta_{min}$, and it decreased by about 1.5\% with each increase of 1\% in $\dot\theta_{min}$.  When the new side wall with 24 thermistors with height $L = 49.54$ cm was used as part of the large apparatus, $\omega_r$ increased by nearly a factor of 2.  It was also found that $p(\theta_0)$  had a smaller peak and was closer to a uniform distribution (i.e. $p(\theta+0) = 1$) with the newer side wall, but other measured parameters such as Reynolds numbers did not change. The side walls were nominally identical, but the diameters had local variations of about 5 parts in 10,000.  It is likely that the first side wall was less circular, resulting in a pressure that tended to force the LSC orientation to align with the long diameter of the side wall.  This could explain why $p(\theta_0)$ had a stronger peak at the preferred orientation in the first side wall, and would suggest that this forcing of the LSC into a preferred orientation also suppreses reorientations.  This suggests an incredible sensitivity of reorientations to the geometry of the system, and it is unclear why it would change the frequency of reorientations by such a large factor. This sensitivity does not seem to qualitatively affect other aspects of reorientations, such as $p(|\Delta\theta|)$.   While we can say that reorientations occur on the order of once per hour, it is difficult to draw more specific conclusions on this subject.

\begin{figure}
\centerline{\psfig{file=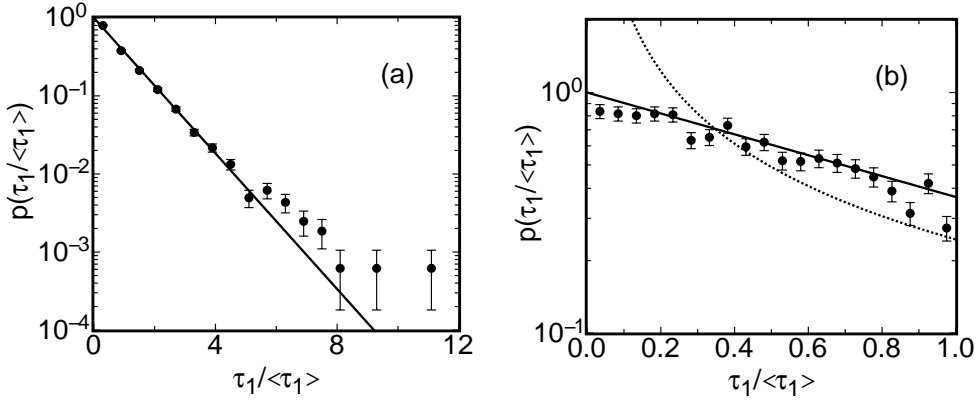,width=5.1in}}
\caption{Solid circles: probability distribution $p(\tau_1 / \langle\tau_1\rangle)$ of the time intervals $\tau_1$ between successive reorientations for all $R$.  (a): Over the entire range of $\tau_1$ measured. (b): For $\tau_1 < \langle\tau_1\rangle$.  Solid lines:  the function $p(\tau_1/\langle\tau_1\rangle) = \exp(-\tau_1/\langle\tau_1\rangle)$, representing the Poisson distribution.  Dotted line:  fit of the power law $p(\tau_1/\langle\tau_1\rangle) \propto (\tau_1/\langle\tau_1\rangle)^{-1}$ to the data, indicating a poor fit.}
\label{fig:prob_time_intervals}
\end{figure}

We now consider the distribution of reorientations in time.  Let $\tau_n$ be the time intervals between the $i$th and $(i+n)$th reorientations.  For $n=1$, $\tau_1$ is simply the time interval between successive reorientations. This is similar to the definition used by \cite{NSSD01}, whose studies of the time intervals between events referred to by them as reversals of the LSC will be compared with the present work.  However, their reversals were defined to occur at one instant, while in the present work reorientations are defined to have some duration. Thus we define $\tau_1$ as the time between, but not including the duration of, reorientations.  The results are essentially the same when the analysis is done with the duration of reorientations included in $\tau_1$.

All of the time intervals $\tau_1$ were sorted into bins according to the value of $\tau_1/\langle\tau_1\rangle$, where $\langle ... \rangle$ represents an average over a data set at a single value of $R$.  The probability distribution  $p(\tau_1 / \langle\tau_1\rangle)$ is  shown in Fig.~\ref{fig:prob_time_intervals}a over the full range of $\tau_1/\langle\tau_1\rangle$, and in Fig.~\ref{fig:prob_time_intervals}b over the limited range $\tau_1 < \langle\tau_1\rangle$.  The error bars indicate the probable error of the mean for each bin.  The data are in good agreement with the exponential function $p(\tau_1/\langle\tau_1\rangle) = \exp(-\tau_1/\langle\tau_1\rangle)$, which represents the Poissonian distribution. Note that there are no adjustable parameters in this comparison.  When reorientations follow Poissonian statistics, it means that successive reorientations occur independently of each other.

The agreement of an exponential function with $p(\tau_1)$ for their reversals was found also by \cite{SBN02}, but only for large $\tau_1$.  They fit a power-law distribution $p(\tau_1) \propto \tau_1^{-1}$ to their data for small time intervals, with $\tau_1 \stackrel{<}{_\sim} 1000 s \simeq 30 {\cal T}$. This time interval roughly corresponds to $\tau_1 < \langle\tau_1\rangle$ in the present work.  A fit of the function $p(\tau_1/\langle\tau_1\rangle) \propto (\tau_1/\langle\tau_1\rangle)^{-1}$  to our data is shown as a dotted line in Fig.~\ref{fig:prob_time_intervals}b for comparison; it is not a good representation of our data. For our results the exponential distribution $p(\tau_1/\langle\tau_1\rangle) = \exp(-\tau_1/\langle\tau_1\rangle)$ (solid line) continues to hold also at small $\tau_1$. The cause of this difference between our results and those of \cite{SBN02} will be discussed further in Sect.~\ref{sec:comparison}.

 \begin{figure}                                                
\centerline{\psfig{file=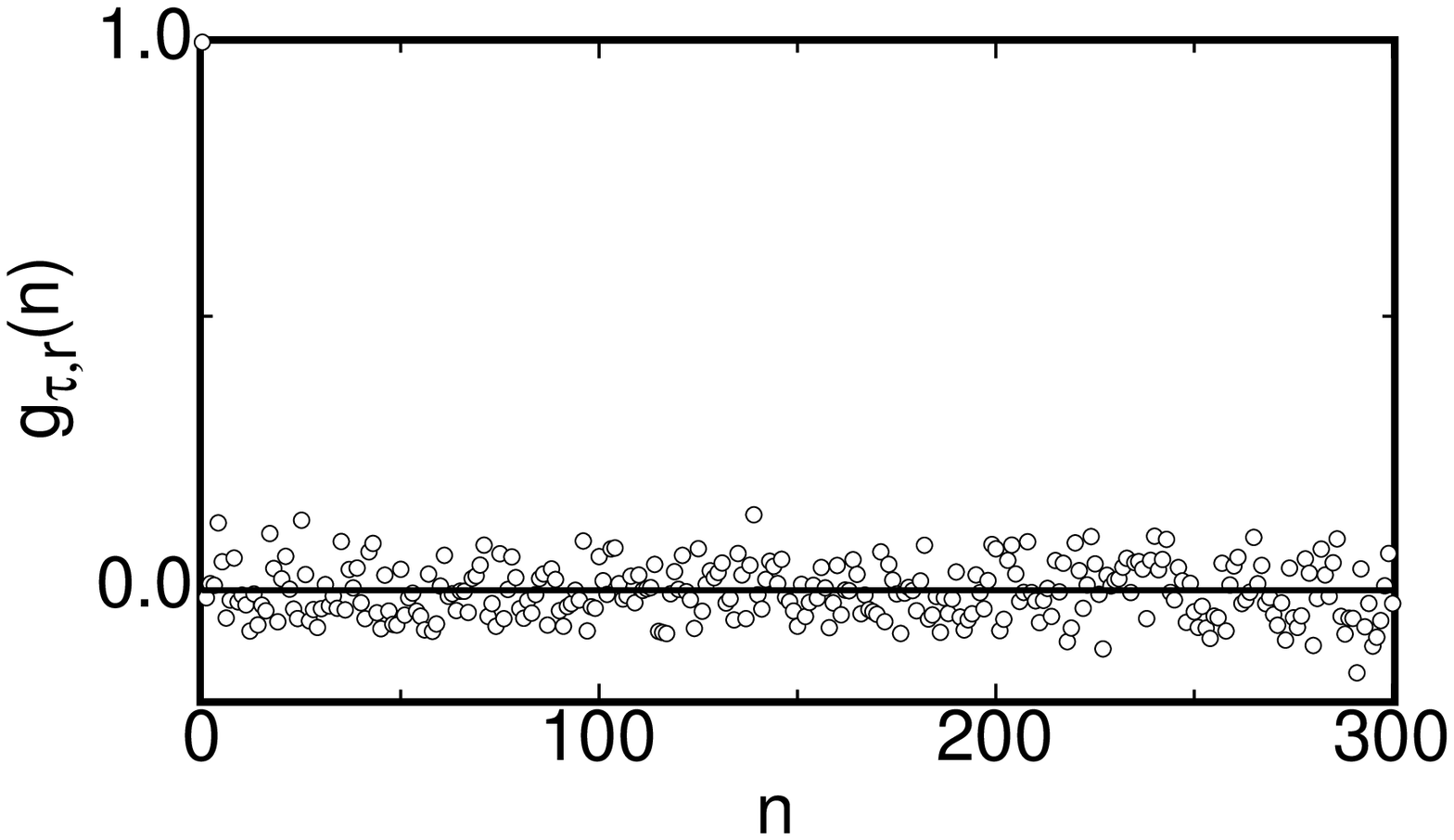,width=3.5in}}
 \caption{The autocorrelation of time intervals between reorientations $g_{\tau,r}(n)$ for $R=1.1\times10^{10}$ in the medium sample. A Poissonian process should yield a delta function at $n=0$ and is consistent with the data. }
 \label{fig:autocorr_time}                                      
 \end{figure}

As another test of the Poissonian nature of reorientations, an autocorrelation of successive time intervals  is given by:

\begin{equation}
g_{\tau,r}(n) =\frac{ \langle(\tau_{1,k+n}-\langle\tau_1\rangle)(\tau_{1,k}-\langle\tau_1\rangle)\rangle }{ \langle(\tau_{1,k}-\langle\tau_1\rangle)^2\rangle}\ .
\label{eq:g_tau}
\end{equation}

\noindent Here $\tau_{1,k}$ is the $k$th time interval between successive reorientations when they are arranged in order of occurrence.  The plot of $g_{\tau,r}(n)$ is shown in Fig.~\ref{fig:autocorr_time} for $R=1.1\times 10^{10}$.   According to the normalization, $g_{\tau,r}(0) = 1$  but for all other $n > 0$, $g_{\tau,r}(n)$ is scattered around zero.  The autocorrelation function for a perfect Poisson process is a delta function, while a finite sample size would result in some scatter around zero.  Figure~\ref{fig:autocorr_time} shows good agreement with Poissonian statistics.

 \section{Cessation statistics}
 \label{sec:cessation}

During the entire investigation, spanning about one year of data acquisition in each of the two samples,  we observed a total of nearly 1000 cessations. Of these 694 were for untilted samples and at the Prandtl number $\sigma = 4.38$ under consideration in this paper. In addition, 52 cessation events were encountered for  $\sigma = 4.38$ in samples that were tilted relative to gravity at various angles. 
 
\subsection{Definition of cessations} 
 
To better distinguish between the rotation and the cessation mechanism, we now consider statistics for cessations only. In a previous paper [\cite{BNA05a}] we identified cessations as a subset of reorientations by determining  whether the amplitude $\delta$ had dropped below a specified value during the reorientation. We found that cessations accounted for about 5\% of reorientations.  However,  there also were events where the amplitude dropped without a significant change in orientation.  These should be counted as cessations, but were not counted as reorientations. Thus we now redefine cessations and count them whenever $\delta$ drops below a chosen minimum amplitude $\delta_{l}$.  All of the adjacent points in the time series are counted as part of the cessation as long as $\delta$ is below a chosen maximum amplitude $\delta_h > \delta_l$.    The lower-amplitude threshold $\delta_{l}$ was chosen as the largest value such that $p(|\Delta\theta|)$ was uniform.  It was found to depend on the side wall used, but for each side wall it was chosen as the largest value such that $p(|\Delta\theta|)$ was uniform. For all side walls this limiting distribution of uniform $p(|\Delta\theta|)$ was reached for some small $\delta_{l}$ and $p(|\Delta\theta|)$ remained uniform in the limit as $\delta_{l}$ was reduced to zero.  For the medium sample, and for the large sample side wall with eight thermistors, we used the parameter value $\delta_{l} = 0.15\langle\delta\rangle$.  For the large sample side wall with 24 thermsitors, we used $\delta_l = 0.07\langle\delta\rangle$.  The upper-amplitude threshold $\delta_{h}$ was chosen as the largest value such that $\dot\delta$ was equal to its limiting value for small $\delta$.  For all samples this was  $\delta_{h} = 0.5\langle\delta\rangle$.  These properties of cessations will be explained in more detail later in this section.   Both parameters depend on the average amplitude $\langle\delta\rangle$ at each $R$, which  increases with $R$.  Because our definition of cessations depends on $\delta$ while reorientations depended on $\Delta\theta$ and $\dot\theta$, some events fall into both categories, possibly with different starting and ending times.  Other events are unique to either reorientations or cessations.

  \begin{figure}
\centerline{\psfig{file=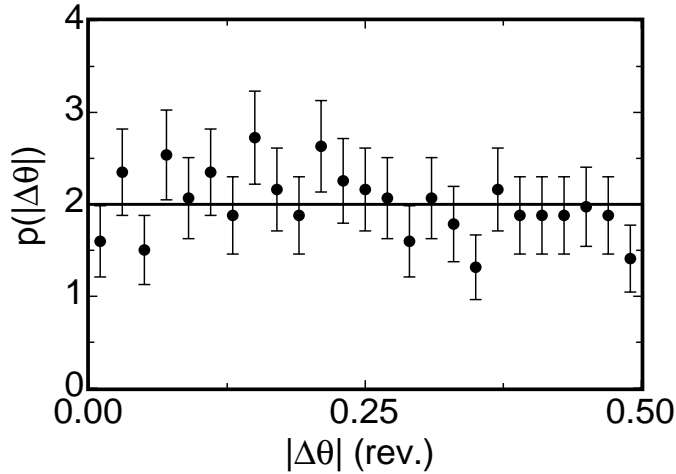,width=3.5in}}
\caption{The probability distribution $p(|\Delta\theta|)$ of the net angular change during cessations for all $R$.   Solid line:  the uniform distribution.}
\label{fig:cess_delta_theta}
\end{figure}

\subsection{Results} 

 The probability distribution $p(|\Delta\theta|)$ of the net angular change during cessations is shown in Fig. \ref{fig:cess_delta_theta} for data at all $R$ from both samples and including diffferent side walls with different $\delta_l$.  For cessations we calculated $\Delta\theta$ reduced to the range $-\pi < \Delta\theta < \pi$ by adding or subtracting multiples of $2\pi$, and then further reduced it to $|\Delta\theta|$.  The probability distribution is seen to be consistent with the uniform distribution.  This agrees with our earlier results for $p(|\Delta\theta|)$ [\cite{BNA05a}]Œ, where we counted cessations as a subset of reorientations.  This current plot additionally covers the range $0 \le |\Delta\theta| < 1/8$ rev., which could not have been counted using the reorientation algorithm. This $p(|\Delta\theta|)$ for cessations is very different from the distribution for  reorientations, which was found to follow a power law with a large negative exponent.  The uniform distribution of angular changes implies that after the LSC stops, it is equally likely to start up again at any new orientation, apparently losing its memory of its previous orientation.  This interpretation is the reason for using the reduced range $-\pi < \Delta\theta < \pi$, since there is no physical difference between choices of $\Delta\theta$ separated by $2\pi$ for cessations, in contrast to the case for rotations where there is a continous variation in $\theta_0$.  Again, reversals (i.e. $\Delta\theta = 1/2$ rev.) are not especially common.  
 
\begin{figure}
\centerline{\psfig{file=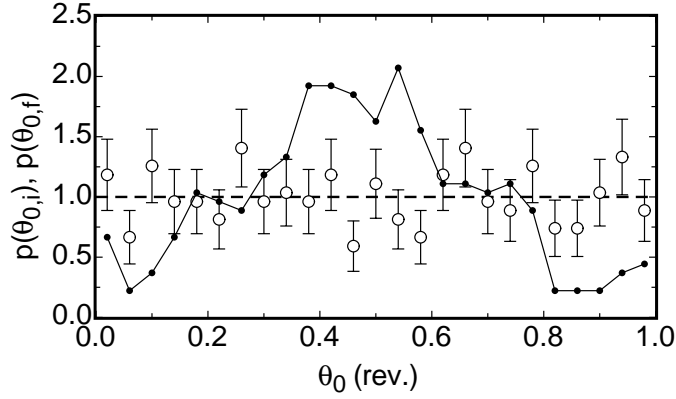,width=3.5in}}
\caption{The probability distributions $p(\theta_{0,i})$ and $p(\theta_{0,f})$ of the orientation at the beginning (solid connected circles, $p(\theta_{0,i})$) and the end (open circles, $p(\theta_{0,f})$) of cessations. Dashed line:  the uniform distribution.  Data is for all $R$ in the medium sample. }
\label{fig:cess_prob_theta}
\end{figure}

Figure \ref{fig:cess_prob_theta} shows the probability distributions $p(\theta_{0,i})$ (solid squares) and $p(\theta_{0,f})$ (open circles) of the orientations $\theta_{0,i}$ and $\theta_{0,f}$  at the beginning and at the end of cessations for all data from the medium sample.  The error bars represent the probable error of the mean of each bin.  The distribution $p(\theta_{0,i})$ at the beginning of cessations is peaked near a preferred orientation $\theta_m$, much like $p(\theta_0)$ (see Fig.~\ref{fig:prob_theta}).  This distribution is non-zero for all $\theta_0$, so cessations can occur at any orientation.  The distribution $p(\theta_{0,f})$ at the end of cessations is consistent with a uniform distribution, thus it is consistent with our conclusion that the LSC restarts at a random orientation after a cessation. It is interesting to note that whatever inhomogeneity causes the maximum in $p(\theta_0)$ and $p(\theta_{0,i})$ does not have any effect on $p(\theta_{0,f})$. This is consistent with the fact that cessations are relatively quick events, happening in about a turnover time, while the non-uniform $p(\theta_0)$ can be attributed to the net effect over a long period from a weak forcing [\cite{BA06}].

   \begin{figure}
\centerline{\psfig{file=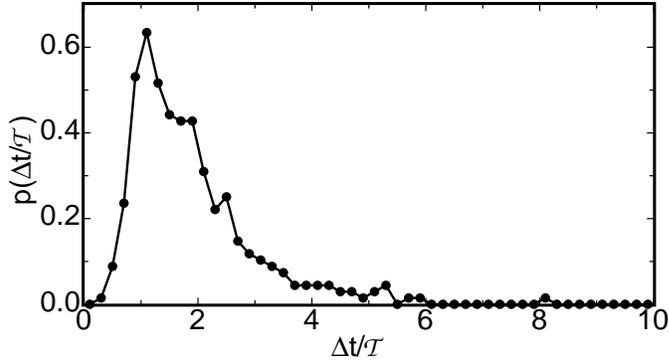,width=3.5in}}
\caption{The probability distribution of the normalized duration of cessations $p(\Delta t/{\cal T})$ for all $R$.}
\label{fig:cess_duration}
\end{figure}
 
One interesting quantity is the duration of cessations $\Delta t$.   Figure \ref{fig:cess_duration} shows the probability distribution $p(\Delta t/{\cal T})$ of $\Delta t$ normalized by the turnover time.  The figure shows a peak near $\Delta t/{\cal T} \simeq 1.2$.  However, this peak location is dependent on the parameters we use in the defintion of cessations, since it represents the time for the amplitude to change from $\delta_h$, down to below $\delta_l$, and back up again to $\delta_h$.  

\begin{figure}
\centerline{\psfig{file=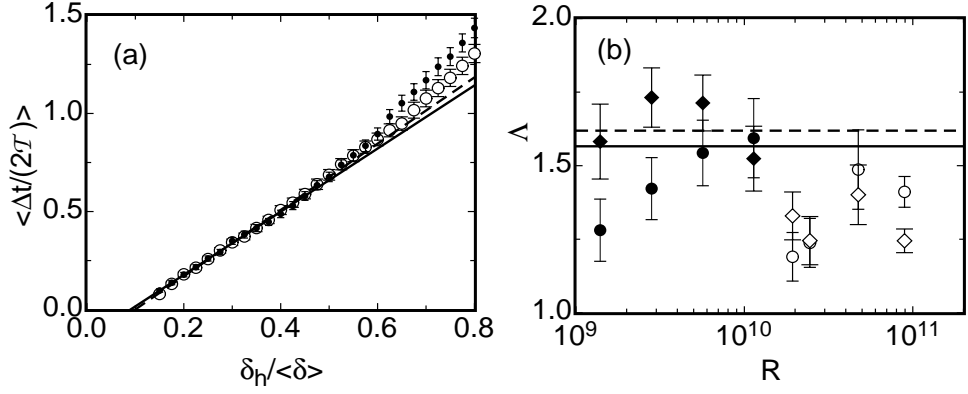,width=5.1in}}
\caption{(a):  The average normalized half-duration of cessations $\langle\Delta t/(2{\cal T})\rangle$ for different values of the cutoff parameter $\delta_h/\langle\delta\rangle$ for all $R$.  Solid circles:  duration of amplitude decrease.  Open circles:  duration of amplitude increase.  Solid line:  fit of the linear function $\langle\Delta t/(2{\cal T})\rangle =\Lambda(\delta_h - \delta_0)/\langle\delta\rangle$ to the data with $\delta_{h}/\delta < 0.5$ for the amplitude decrease.  Dashed line:  fit of the same function to the amplitude increase.  (b):  The fit parameter $\Lambda$ for several values of $R$ in the medium sample (solid symbols) and the large sample (open symbols) for the amplitude decrease (circles) and increase (diamonds).  Solid line:  the average $\Lambda$ over all $R$ for the amplitude decrease.  Dashed line:  the average $\Lambda$ over all $R$ for the amplitude increase.}
\label{fig:cess_amp_rate}
\end{figure}

To determine a more physically meaningful value relating to the duration of cessations, the duration of each cessation was calculated for several values of the cessation cutoff parameter $\delta_h$.  Each cessation was also divided into two time intervals, one before and one after the minimum amplitude was reached, so that the durations of the amplitude decrease and the amplitude increase could be determined separately.  Figure \ref{fig:cess_amp_rate}a shows the average cessation half-duration $\langle\Delta t/(2{\cal T})\rangle$ for both the amplitude decrease (solid circles) and the amplitude increase (open circles) for different values of $\delta_h/\langle\delta\rangle$.  Cessations at all $R$ were used so as  to obtain a sufficiently large collection of events.  The linear function $\langle\Delta t/(2{\cal T})\rangle = \Lambda(\delta_h - \delta_0)/\langle\delta\rangle$ was fit to the data for $\delta_h/\langle\delta\rangle < 0.5$. This yielded $\Lambda = 1.61 \pm 0.04$ for the amplitude decrease and $\Lambda = 1.69 \pm 0.04$ for the amplitude increase. It is interesting to note that on average the decay of the LSC during a cessation took just as long as the subsequent growth. The fit also gave $\delta_0/\langle\delta\rangle = 0.090 \pm 0.005$ for the amplitude drop and $\delta_0/\langle\delta\rangle = 0.099 \pm 0.004$ for the amplitude rise.  The value of $\delta_0$ was close to the average minimum amplitude for cessations $\langle \min(\delta)/\langle\delta\rangle\rangle = 0.088 \pm 0.003$, and probably represents a base level of temperature fluctuations without the LSC.  

Inverting the fitting function yields a characteristic rate of change of the amplitude $|\dot\delta| = 2(\delta_h-\delta_0)/\Delta t = 1/\Lambda\times\langle\delta\rangle/{\cal T} = 0.62\pm 0.01 \langle\delta\rangle/{\cal T}$ for the amplitude decrease and $0.59\pm 0.01 \langle\delta\rangle/{\cal T}$ for the amplitude increase.  Figure~\ref{fig:cess_amp_rate}b shows how $\Lambda$ varies with $R$ for several data sets with at least 14 cessations each. This shows that the slope with the normalizations used above is independent of $R$, within the precision of the experiment.  This justifies the use of data with all values of $R$ in Fig.~\ref{fig:cess_amp_rate}a.  It is interesting that $\dot\delta$ has the same magnitude during both the decrease in amplitude and increase in amplitude, and remains constant for much of the duration of cessations.  We had no reason to expect this, and this should put a significant restriction on any dynamical theories of cessations.  The value of $\dot\delta$ also implies a physical time scale for cessations, so for a cessation with the amplitude starting and ending at $0.5\langle\delta\rangle$, the duration is on average $\Delta t = 2(\delta_h-\delta_0)/|\dot\delta| \approx 1.2 {\cal T}$.  

\begin{figure}
\centerline{\psfig{file=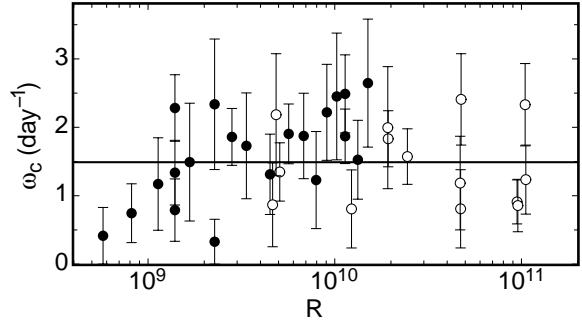,width=3.0in}}
\caption{The average rate of occurrence of cessations $\omega_c$ versus $R$.  Solid circles:  medium sample.  Open circles:  large sample.  Solid line:  average frequency of cessations for all data.}
\label{fig:cess_rate}
\end{figure}
 
The average rate of occurrence of cessations $\omega_c$ versus $R$ is shown in Fig.~\ref{fig:cess_rate} for data sets that are at least two days long. The error bars represent the probable error of the mean for each data set. The data are consistent with a rate $\omega_c$ independent of $R$ over the range studied, except for a decrease in $\omega_c$ for the smallest $R$.  Over 226 days of total running time, the average rate was $1.49 \pm 0.08$ day$^{-1}$.  It should also be noted that $\omega_c$ does depend on the cessation definition parameter $\delta_l$, which varied with the side wall used, although  curiosly $\omega_c$ did not change significantly with the side wall used.  With each side wall considered separately, there is about a 1.7\% increase in the cessation count with each 1\% increase in $\delta_l/\langle\delta\rangle$, although the value of $\delta_l$ chosen is significant because it is the largest value that leads to a uniform $p(\Delta\theta)$.
 
\begin{figure}
\centerline{\psfig{file=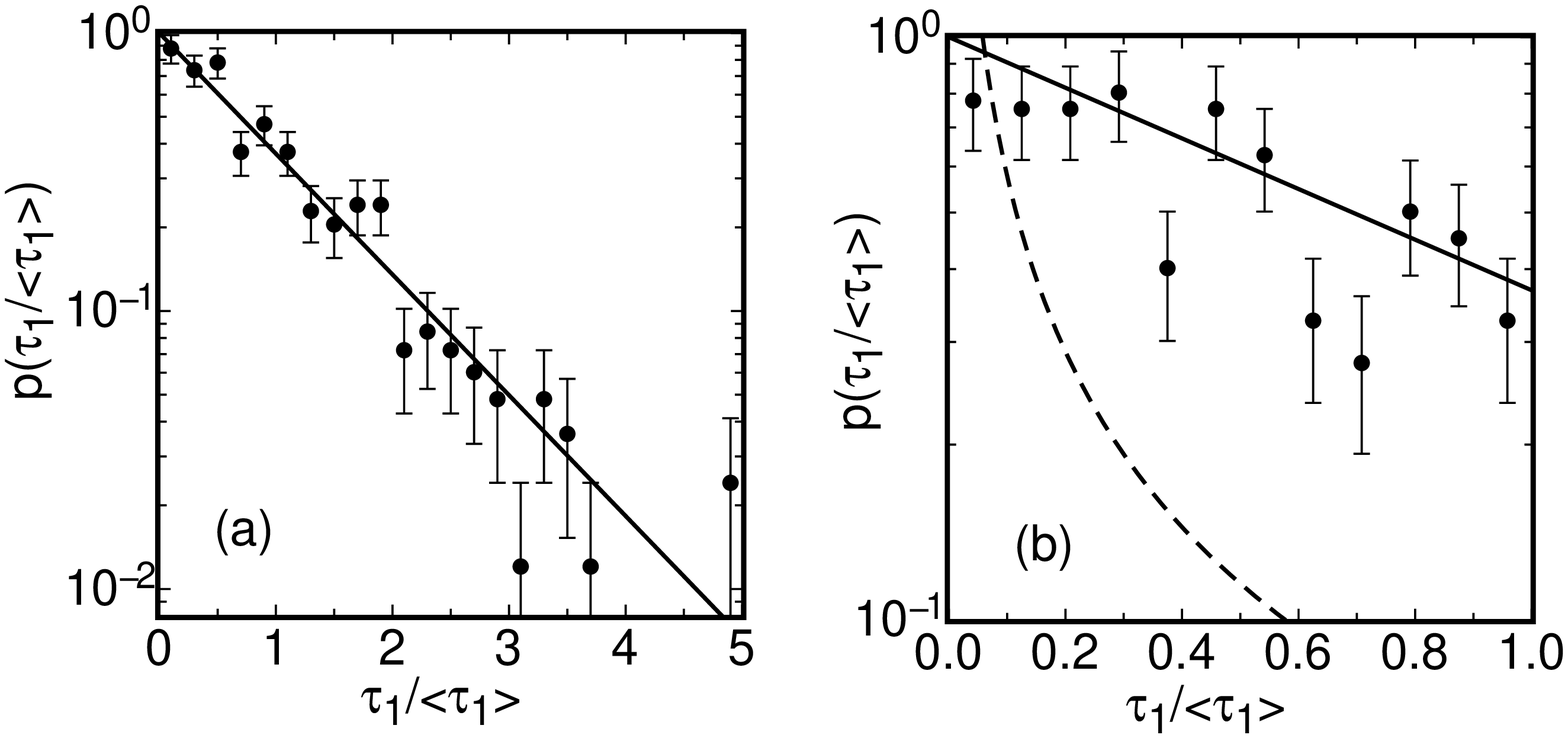,width=5.1in}}
\caption{The probability distribution of time intervals between cessations $p(\tau_1/\langle\tau_1\rangle)$ for all $R$. The entire range of the data is shown in (a), while a range restricted to $\tau_1 < \langle\tau_1\rangle$ is shown in (b) for better resolution. Solid lines:  $p(\tau_1/\langle\tau_1\rangle) = \exp(-\tau_1/\langle\tau_1\rangle)$, representing the Poisson distribution.  Dashed line:  power law with an exponent of $-1$, showing a poor fit to the data for $\tau_1 < \langle\tau_1\rangle$. }
\label{fig:cess_time_interval}
\end{figure}
 
The probability distribution of time intervals between cessations $p(\tau_1/\langle\tau_1\rangle)$ for all data from the medium sample is shown in Fig.~\ref{fig:cess_time_interval}a, with error bars representing the probable error of the mean.   The exponential function $p(\tau_1/\langle\tau_1\rangle) = \exp(-\tau_1/\langle\tau_1\rangle)$ agrees well with the data, indicating that cessations follow Poissonian statistics in time, as did reorientations.  Figure \ref{fig:cess_time_interval}b shows $p(\tau_1/\langle\tau_1\rangle)$ over the limited range $\tau_1< \langle\tau_1\rangle $, which shows that the data are consistent with the same exponential function (solid line) in this region.  This distribution can not be fitted by a  power law with exponent $-1$ (dashed line), as found by \cite{SBN02} for their reversals. This difference between the two experiments will be examined further in the next Section.

\begin{figure}
\centerline{\psfig{file=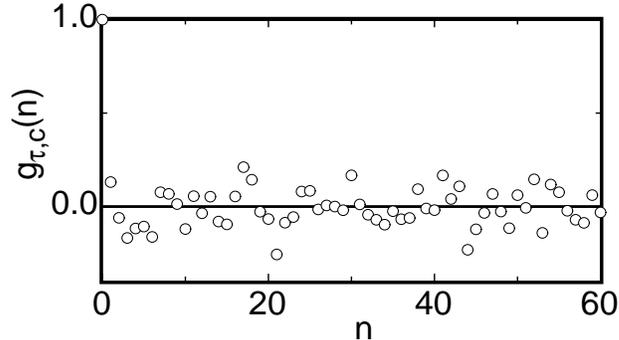,width=3.2in}}
\caption{The autocorrelation of time intervals between successive cessations $g_{\tau,c}(n)$ for $R=1.1\times 10^{10}$ in the medium sample. A Poissonian process should yield a delta function at $n=0$ and is consistent with the data. }
\label{fig:cess_autocorr_time}
\end{figure}

The autocorrelation of the time intervals between successive cessations $g_{\tau,c}(n)$, defined in a manner analogous to that given by Eq.~\ref{eq:g_tau},  is plotted in Fig.~\ref{fig:cess_autocorr_time} for $R = 1.1\times 10^{10}$.  This data is from a run that contains 116 cessations over 54 days.  Much like for reorientations, we find only scatter about $g_{\tau,c}(n) = 0$ for $n > 0$, providing more evidence that cessations have Poissonian statistics in time.

\section{ Comparison of temporal statistics with earlier experiments}
\label{sec:comparison}

An extensive study of LSC ``reversals" was reported by \cite{NSSD01}. These authors used a $\Gamma \simeq 1$ sample filled with helium gas at temperatures near 5 Kelvin. They reported data from a single pair of temperature sensors located at half-height in the fluid near the side wall. At that location they could determine the prevailing vertical component of the velocity of hot or cold temperature fluctuations, i.e. of ``plumes". It is generally held that the plume velocity is the same as that of the LSC.  They defined reversals of the LSC as any switch between up-flow and down-flow of this local vertical velocity component.

In a sequence of subsequent papers [\cite{SBN02,NS02,NSSD02,SBN04,HYBNS05}] some of the members of the same research group carried out various statistical analyses of a particular data set taken at a Rayleigh number of $1.5\times 10^{11}$, i.e. close to the highest Rayleigh number achieved in the present work . For this time series the Prandtl number was 0.74 [\cite{NS03}], which differs somewhat from our $\sigma = 4.38$. Although the value of $\sigma$ may have some influence for instance on the frequency of reversals, it seems unlikely that the difference in $\sigma$ would qualitatively alter the physics of the reversals. The turnover time in our experiments is 49 s at $R=1.1\times10^{10}$, while \cite{NSSD01} report a turnover time of about 30 s for $R=1.5\times10^{10}$, so if the characteristic time scales for reversals are proportional to  the turnover time, then we should expect them to be a little shorter for \cite{NSSD01}.

The major difference between the results of their work and of ours is the distribution of $p(\tau_1/\langle\tau_1\rangle)$ for the time intervals $\tau_1$ between successive events.  \cite{SBN02} found that a power law fit the distribution $p(\tau_1/\langle\tau_1\rangle)$ for small $\tau_1$ and an exponential function fit the tail, while we found an exponential function to fit the distribution over the entire range for both reorientations and cessations (see Figs.~\ref{fig:prob_time_intervals} and \ref{fig:cess_time_interval}).  In a related matter there is a large difference between the frequency of events determined in the two experiments:  \cite{SBN02} reported about 420 events per day while we find only about 50 reorientations per day -- an order of magnitude less.  If their reversals correspond to cessations, then the difference is 2 orders of magnitude; or if the correspondence is to reorientations, for instance with $\Delta\theta = 1/2 \pm 0.05$ rev., then the difference is 3 orders of magnitude.

These differences can be understood by reinterpreting some results.  \cite{NSSD01} reported a nearly bimodal velocity distribution, which led them to the then reasonable belief that the large-scale circulation-plane had a more or less fixed azimuthal orientation relative to their sensors, and that the LSC was switching between two opposing directions in that plane.  Our experiments with eight side-wall thermistors provided new information in the azimuthal dimension, showing that the LSC in our system samples all azimuthal orientations, and that reorientations and cessations can occur at any orientation.   While it is possible that an asymmetry of the apparatus used by \cite{NSSD01} could have resulted in two opposing preferred orientations, it seems unlikely that this could also account for the much greater frequency of events observed by them, especially since our experiments with an asymmetry introduced by tilting the samples (see Sect.~\ref{sec:tilt}) show a strongly {\it reduced} frequency for both reorientations and cessations, but otherwise no qualitative difference.

Here we present an alternative explanation for the results of \cite{NSSD01}.  We consider the velocity in the plume layer, a thin region between the bulk and boundary layer near the side wall. We hypothesize that this velocity, when averaged over an intermediate time interval (say of order ${\cal T}$), has a vertical component that is nearly a step function of $\theta$, with speed $+v$ when $\theta$ is within $\pm \pi/2$ rad of $\theta_0$ and $-v$ for all other angles.  To be clear, this time average should be over a period long enough to identify the LSC, but short enough to not contain any reorientations.  The one-turnover-time average used in the experiment of \cite{NSSD01} satisfies these criteria.  A justification for this step-function velocity-distribution, independent of its ability to explain the results of \cite{NSSD01}, is given in the Appendix.  As a consequence of random azimuthal meandering of the LSC this distribution would produce the bimodal velocity distribution reported by \cite{NSSD01}. It also leads to an alternative interpretation of the events they counted:  if reversals are counted whenever the local velocity switches directions at a sensor as seen by \cite{NSSD01}, then this is equivalent to counting events whenever the LSC orientation {\it crosses} a  critical angle orthogonal to the azimuthal sensor orientation so that the interface between the up- and down-flow moves past the sensors.  These events usually are not reversals with $\Delta\theta = 1/2$ rev., as they would include events covering a continuum of $\Delta \theta$. Indeed for many of the events the LSC orientation would change only slightly, but just enough to cross the critical  angle.  We can test this hypothesis by reanalyzing our own data to study this new type of  event which we shall call a {\em crossing}. In our system we shall count it whenever the LSC orientation $\theta_0(t)$ crosses either of the two angles orthogonal to the preferred orientation, i.~e.~$\theta_m \pm \pi/2$ rad.  This choice of the crossing angles is arbitrary as long as they are separated by $\pi$, and if we use for instance $\theta_m$ and $\theta_m + \pi$ rad then our statistical results are qualitatively unchanged.  If our hypothesis about the velocity distribution is correct, then the analysis of crossings from our data should yield the same statistical results as the events reported by \cite{NSSD01} and \cite{SBN02}.  

\begin{figure}
\centerline{\psfig{file=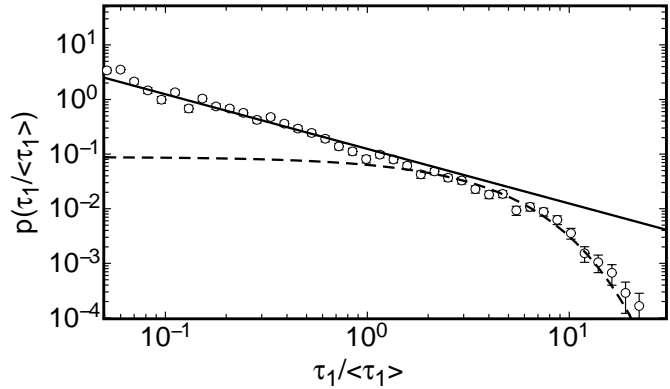,width=3.5in}}
\caption{$p(\tau_1 / \langle\tau_1\rangle)$ for crossings for $R= 1.1\times 10^{10}$ in the medium sample, defined as occurring when the orientation of the LSC crosses either angle orthogonal to the preferred orientation $\theta_m$.  Solid line:  a fit of the power law $p(\tau_1 / \langle\tau_1\rangle) \propto \tau_1 / \langle\tau_1\rangle^{-1}$ to the data for $\tau_1 < \langle\tau_1\rangle$.  Dashed line:  a  fit of an exponential function to the data for $\tau_1  >  \langle\tau_1\rangle$.}
\label{fig:crossings_rev}
\end{figure}

The probability distribution of the time intervals $p(\tau_1/\langle\tau_1\rangle)$ for crossings is shown in Fig.~\ref{fig:crossings_rev} at $R=1.1\times 10^{10}$, with error bars equal to the probable error of the mean as before.  For one data set we found 2032 crossings over 11.8 days (for comparison we found 555 reorientations and 21 cessations in the same data set), corresponding to  $\langle\tau_1\rangle = 502$ s or about 172 crossings per day. This is  only a factor of 2.4 smaller than the frequency of events reported by \cite{SBN02}, and thus plausibly consistent, especially since their experiment has a somewhat shorter turnover time.

The figure shows a fit of an exponential function $p(\tau_1) \propto\exp[-\tau_1/(h\langle\tau_1\rangle)])$ to the data for $\tau_1 > \langle\tau_1\rangle$ (dashed line), which fits well in that range of $\tau_1$ but falls well below the data for smaller $\tau_1$.  We obtain $h\langle\tau_1\rangle=(3.3 \pm 0.1) \langle\tau_1\rangle= 1600$ s from this fit.  In the other limit, a power law $p(\tau_1 / \langle\tau_1\rangle) \propto (\tau_1 / \langle\tau_1\rangle)^{-1}$ was fit to the data for $\tau_1 < \langle\tau_1\rangle$ (solid line), which fits well in that range of $\tau_1$ but is much higher than the data for larger $\tau_1$.  These results of both the frequency of events and of $p(\tau_1)$ for crossings are in good agreement with the results of \cite{SBN02}, which implies that the two experiments are consistent, but that crossings rather than rotations or cessations are comparable to the events reported in the earlier work.

\begin{figure}
\centerline{\psfig{file=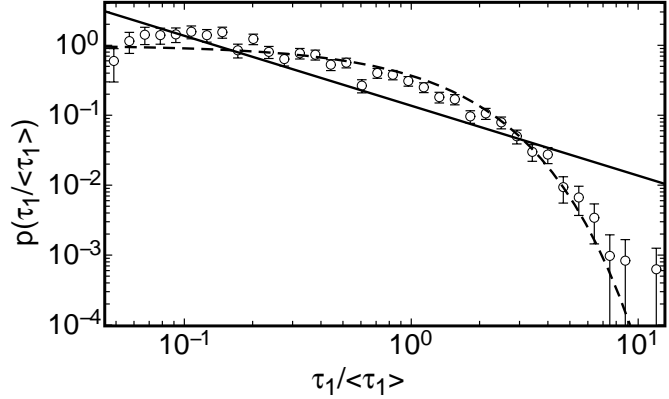,width=3.5in}}
\caption{$p(\tau_1 / \langle\tau_1\rangle)$ for crossings at $R=1.1\times 10^{10}$ in the medium sample, but now there is an extra buffer region of width $\pi/4$ rad around each crossing angle that the orientation must also exit before a crossing is counted. Solid line:  a power law fit of $p(\tau_1 / \langle\tau_1\rangle) \propto (\tau_1 / \langle\tau_1\rangle)^{-1}$ to the data for $\tau_1 / \langle\tau_1\rangle < 1$.  Dashed line:  the exponential function of $p(\tau_1 / \langle\tau_1\rangle) = \exp(-\tau_1 / \langle\tau_1\rangle)$, representing the Poisson distribution.}
\label{fig:crossings_rev_buffer}
\end{figure}

To understand why the power-law distributions occur, we again reanalyze the present data with a modified definition for crossings. It was suggested by \cite{SBN02} that the LSC orientation undergoes some ``azimuthal drift (or jitter)". We want to eliminate the counting of events  in which the orientation jitters back and forth around the crossing angles, and only count events where there is a significant change in orientation of the LSC.  To this avail we define buffered crossings as occurring when $\theta_0(t)$ not only crosses $\theta_m \pm \pi/2$ rad, but also exits a buffer region of width $\pi/4$ rad centered about either crossing angle.  The probability distribution $p(\tau_1/\langle\tau_1\rangle)$ for buffered crossings is shown in Fig.~\ref{fig:crossings_rev_buffer} for $R = 1.1\times 10^{10}$.  An exponential function $p(\tau_1) = \exp(-\tau_1/\langle\tau_1\rangle)$ is consistent with the data (dashed line), showing that these buffered crossings are Poissonian.  A power law $p(\tau_1) \propto \tau_1^{-1}$ is also shown (solid line) for comparison.   Now it is apparent that jitter, or small orientation changes of the LSC around the crossing angles, was responsible for the power-law distribution of $p(\tau_1)$ for small $\tau_1$.  There are only 520 buffered crossings from the same data set that had 2032 crossings, so the jitter is of course also the reason for the large number of events.

\begin{figure}
\centerline{\psfig{file=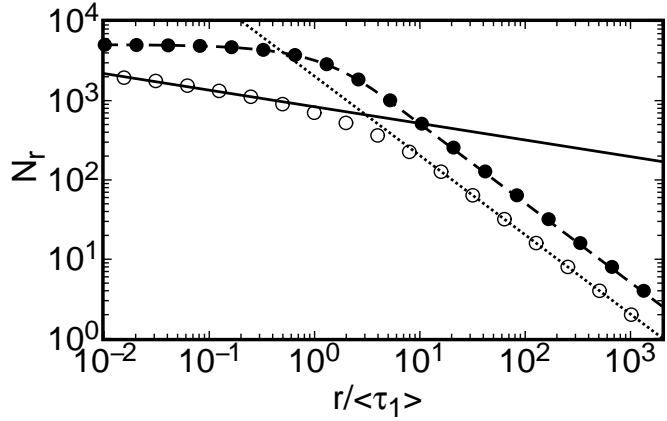,width=3.5in}}
\caption{Number of non-empty bins $N_r$ from the binning of events on the time axis with bin width $r$ for $R=1.1\times 10^{10}$ in the medium sample.  Solid circles:  reorientations. Open circles: crossings. Dashed line: prediction for a Poisson process.   Solid line: power law fit of $N_r = r^{-D}$ to crossings for $r < 0.3$, yielding the fractal dimension $D$.  Dotted line:  space-filling limit $N_r \propto r^{-1}$ for crossings.}
\label{fig:frac_dim}
\end{figure}

We can continue the analysis of the temporal statistics for both reorientations and crossings in the spirit of \cite{NSSD01} and \cite{SBN02}.  One aspect of that analysis was the fractal dimension of the time series of events.  We took a time series with reorientations, divided up the time axis into bins of width $r$,  and then placed each reorientation into the appropriate bin based on the time when the event occurred. We then counted the number of non-empty bins $N_r$. The fractal dimension $D$ is defined by the equation $N_r = r^{-D}$ in the limit of small $r$. Figure \ref{fig:frac_dim} shows $N_r$ for various bin widths $r$ for reorientations at all $R$ (solid circles) and for crossings at $R=1.1\times 10^{10}$ (open circles).  For $r \gg \langle\tau_1\rangle$, the data always follows $N_r \propto r^{-1}$, which is the case where the bin width is so large that every bin contains at least one reorientation. This is called the space-filling limit.  For $r \ll \langle\tau_1\rangle$, $N_r$ reaches a constant ($D=0$) for reorientations.  This is the limit where the bin widths are so small that every reorientation falls into its own bin.  Also shown in Fig.~\ref{fig:frac_dim} is the analytical result for a Poissonian distribution:  $N_r = (N_0\langle\tau_1\rangle/r) \times[1-\exp(-r/\langle\tau_1\rangle)]$, where $N_0$ is the total number of events.  The theory agrees perfectly with our data for reorientations, so it provides further evidence that reorientations are Poissonian.  However, the data for crossings can be fitted by a power law $N_r \propto r^{-D}$ for $r < \langle\tau_1\rangle$, and yield $D=0.21$.  This is reasonably consistent with the results of \cite{NSSD01}, which gave a range of values  $0.2 \le D \le 0.5$ for different $R$.  The non-zero value of $D$ is characteristic of the power-law distribution of time intervals;  no matter how small $r$ is, there are a finite number of time intervals that are smaller than $r$, so $N_r$ will not reach the total number of events.

When we calculate the fractal dimension for buffered crossings we get $D = 0$.  $D$ is indistinguishable from zero for buffer widths greater than $3\pi/16$ rad., while $D$ gradually increases for smaller buffer widths up to $D = 0.21$ for no buffer.   This suggests that once the orientation is $3\pi/32$ rad. from the crossing angle it is far enough away that the jitter of $\theta_0$ no longer affects the event statistics. 

 \begin{figure}                                                
\centerline{\psfig{file=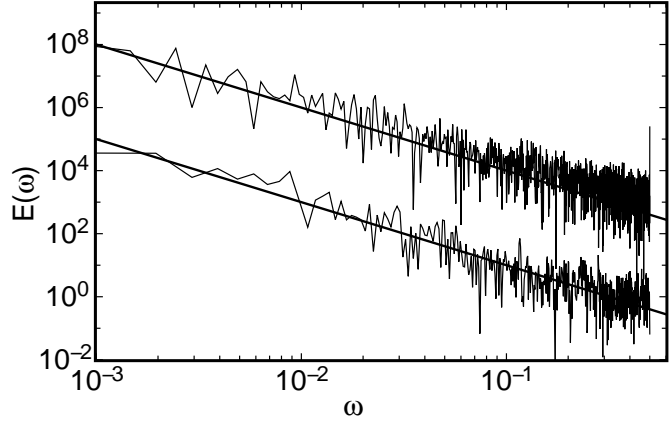,width=3.5in}}
 \caption{The power spectral density $E(\omega)$ of the detrended time $\Delta(k)$ of the $k^{th}$ event for crossings (upper data) and reorientations (lower data) at $R=1.1\times 10^{10}$ in the medium sample.  $E(\omega)$ for reoreintations is shifted down by a factor of 100.  The solid straight lines are power laws with exponents of $-2$.}
 \label{fig:spec_brownian}                                      
\end{figure}

The power-spectral density of the detrended timing of events was reported by \cite{SBN02}.
Let the time of the $k^{th}$ reorientation $t_k$ from the beginning of a data set be defined as $t_k = \langle\tau_1\rangle[k + \Delta(k)]$.  Here $\Delta(k)$ is the deviation from a linear trend of events in time, where the linear trend is $\langle\tau_1\rangle k$ as defined by \cite{SBN02}.  The power-spectral density $E(\omega)$ of $\Delta(k)$ is given by $E(\omega) = 1/(2\pi)| \sum_k \Delta(i)e^{-i\omega k} |^2$.  $E(\omega)$ is plotted in Fig.~\ref{fig:spec_brownian} for both crossings (upper data) and reorientations (lower data). In both cases, $E(\omega)$ is in good agreement with an $\omega^{-2}$ roll off, consistent with a Brownian process, and consistent with the results of \cite{SBN02}.

 \begin{figure}                                                
\centerline{\psfig{file=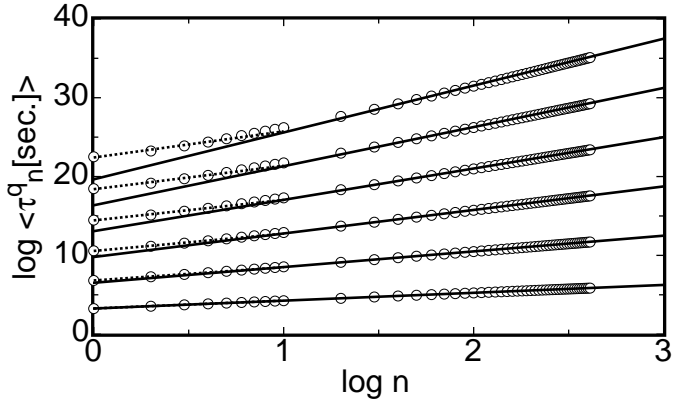,width=3.5in}}
 \caption{Moments $\langle\tau^q_n\rangle$ of the generalized time intervals between reorientations versus $n$th interval for $q = 1, 2, 3, 4, 5, 6$ at $R=1.1\times 10^{10}$ in the medium sample.  Ascending data sets correspond to increasing $q$.  Solid lines:  power law fits for each $q$ of $\langle\tau^q_n\rangle \propto n^{\zeta}$ for $n \ge 100$.  Dotted lines: power law fits for each $q$ of $\langle\tau^q_n\rangle \propto n^{\zeta}$ for $n \le 5$.}
 \label{fig:moments}                                      
 \end{figure}

 \begin{figure}                                                
\centerline{\psfig{file=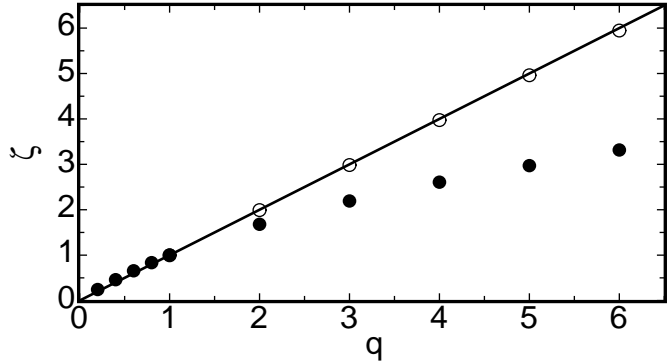,width=3.5in}}
 \caption{Exponents $\zeta$ vs. $q$ from the fits $\langle\tau_n^q\rangle \propto n^{\zeta}$ for $n \le 5$ (solid circles) and for $n \ge 100$ (open circles) for $R=1.1\times 10^{10}$ in the medium sample.  The solid line shows the prediction for Poissonian data.}
 \label{fig:moments_fit}                                      
 \end{figure}
 
The moments of the generalized time intervals between events are given by $\langle\tau^q_n\rangle = \langle(t_{k+n} - t_k)^q \rangle_k$.  The first six moments $1 \le q \le 6$ are shown in Fig. \ref{fig:moments} for reorientations for $R=1.1\times10^{10}$ in the medium sample.   These data can be fitted by power laws $\langle\tau^q_n\rangle  \propto n^{\zeta}$ in limited ranges of $n$ with different $\zeta$ for the large and small limits of $n$.  The $\zeta$ values are shown as a function of $q$ in Fig.~\ref{fig:moments_fit}. The small-$n$ exponents (solid circles) were obtained from fits in the range $n \le 5$, while the large-$n$ exponents (open circles) were from the range $n \ge 100$.  For Poissonian data, we would expect $\zeta = q$ for all $n$.  This is in agreement with the large-$n$ fits.  The small-$n$ data deviate significantly from the Poissonian prediction for large $q$.  \cite{SBN02} reported a similar deviation for their events from the Poissonian prediction. Together with the exponential cutoff of their power-law result for $p(\tau_1)$, they explained this as a finite-size effect.  That explanation will not work for us because we found $p(\tau_1)$ to agree with an exponential distribution over the entire range of $\tau_1$.  Inspection of $p(\tau_1)$ in Fig.~\ref{fig:prob_time_intervals} shows more large time intervals than would be expected for a Poisson distribution, although only by a few events.  This might account for the large higher-order moments that we found, and thus for the offending values of $\zeta$.  The question remains as to whether reorientations are a Poissonian process and we recorded a lot of large time intervals by chance, or whether it indicates non-Poissonian physics at large time intervals.  There are very few data points with these large time intervals, so we can not distinguish between these two possibilities.

 \begin{figure}                                                
\centerline{\psfig{file=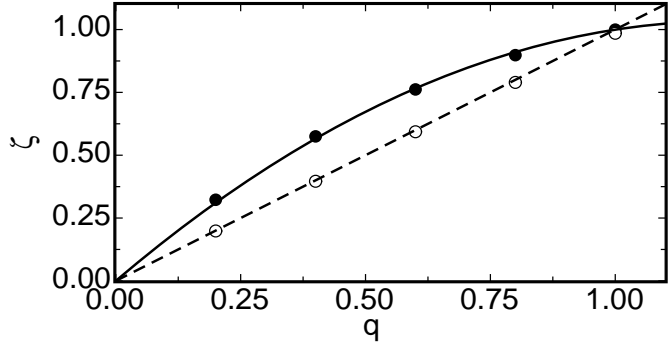,width=3.5in}}
 \caption{Exponents $\zeta$ vs. $q$ from the fits $\langle\tau_n^q\rangle \propto n^{\zeta}$ for crossings at $R=1.1\times 10^{10}$ in the medium sample.  Solid circles:  for $n \le 5$.  Open circles:  for $n \ge 100$.  Dashed line:  linear fit to data for $n \ge 100$. Solid line: quadratic fit to data for $n \le 5$.}
 \label{fig:moments_cross_fit}                                      
 \end{figure}

Again, to better compare to the results of \cite{SBN02}, we calculated the moments for crossings (not shown) and $\zeta(q)$ as before.  Figure \ref{fig:moments_cross_fit} shows the exponents $\zeta(q)$ for both the large-$n$ and small-$n$ limits for crossings, but for $q \le 1$.  For $n \ge 100$, $\zeta \propto q$ as would be expected for Poissonian data.  For $n \le 5$, the quadratic function $\zeta = aq+bq^2$ was fit to the data and yielded $a = 1.69$ and $b = -0.692$.  Similarly, a fit of a quadratic function for $\zeta(q)$ to their data was reported by \cite{SBN02} and attributed to an underlying log-normal distribution.

 \begin{figure}                                                
\centerline{\psfig{file=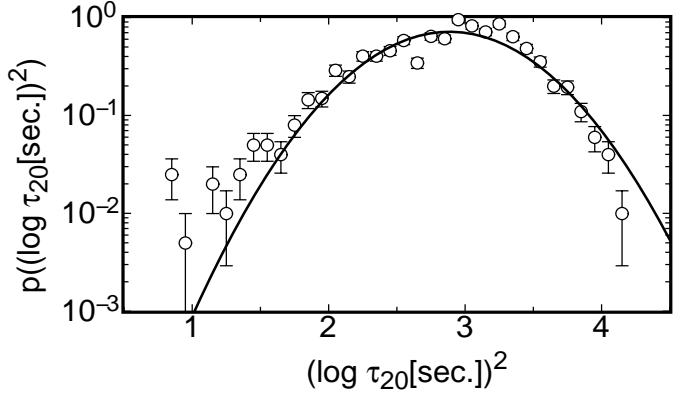,width=3.5in}}
 \caption{The probability distribution of the time intervals between 20 crossings $p(\tau_{20})$ for $R=1.1\times 10^{10}$ in the medium sample.  Solid line:  fit of a bilognormal distribution to the data.}
 \label{fig:bilognormal}                                      
 \end{figure}

Another result from \cite{SBN02} was that $p(\tau_n)$ for an intermediate number of gaps, with $5 < n < 100$, agreed with a bilognormal distribution.  For crossings from the present data, $p([\log\tau_{20}]^2)$ is plotted in Fig. \ref{fig:bilognormal}. A bilognormal function given by
$$p([\log\tau_{20}]^2) \propto \exp\left[ -\frac{[(\log\tau_{20})^2 - m]^2}{\sigma^2} \right]$$
gave a good fit to our data. Again, this shows good agreement between our crossings and the reversals of \cite{SBN02}.

In this section we showed that we can reproduce the major statistical results of \cite{NSSD01} and \cite{SBN02}, but only when we consider crossings. The results from our experiment differ from those of the earlier work when we consider only events corresponding to our definition of reorientations or cessations.   The events reported by \cite{NSSD01} were assumed by them to be reversals of the LSC because the local flow velocity reversed directions.  Our results, based on the angular information about the LSC orientation afforded by the use of eight side-wall thermistors, have shown that they were likely counting  crossings in which the orientation of the LSC crossed a fixed line, which often counts small jitters of the LSC orientation as events.  Since we are interested in events that involve larger changes of the flow field, we have focused on studying reorientations consisting of rotations and cessations.

\section{Azimuthal rotation rate}
 \label{sec:dthetadt} 

\begin{figure}                                                
\centerline{\psfig{file=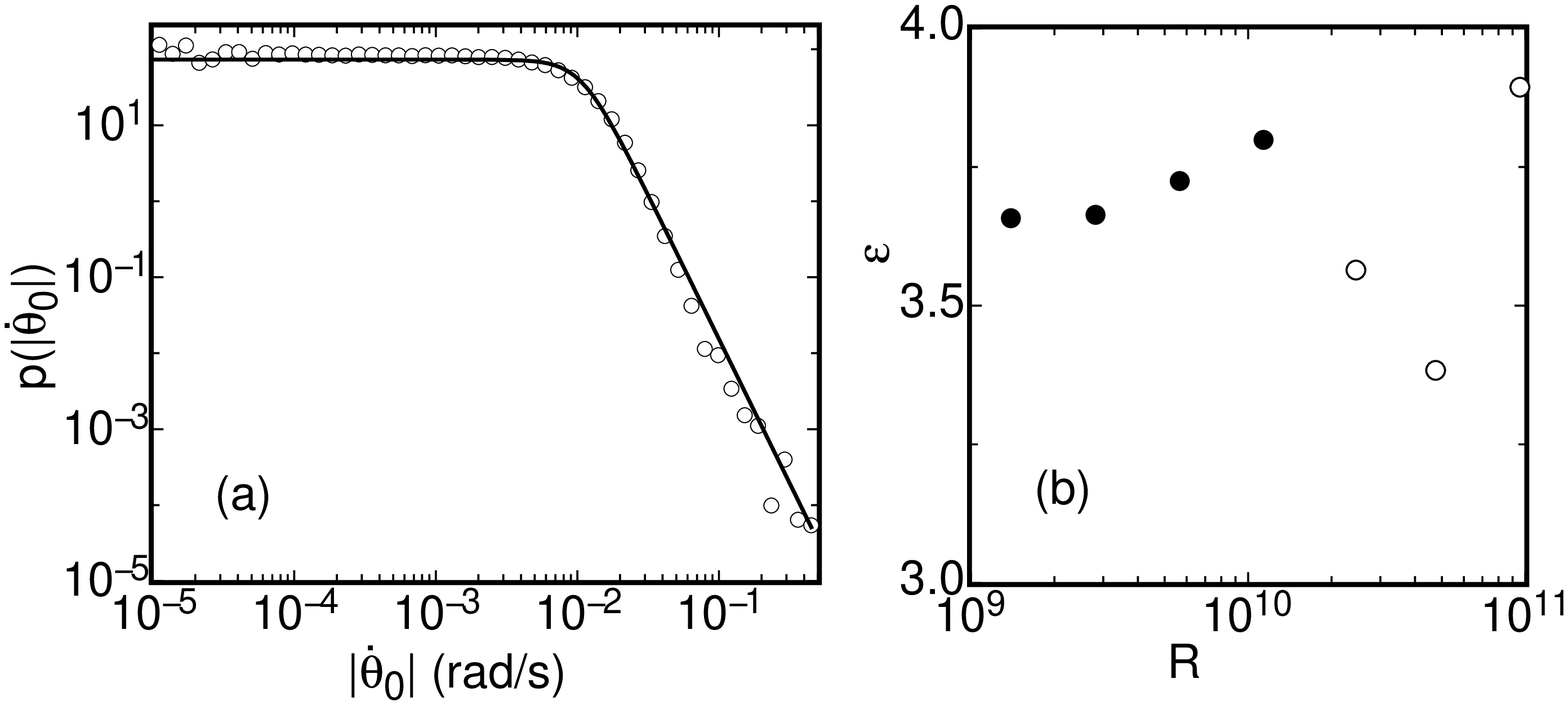,width=5.1in}}
\caption{(a):  The probability distribution of the absolute value of the instantaneous azimuthal rotation rate $p(|\dot\theta_{0}|)$ for $R = 1.1\times 10^{10}$ in the medium sample.  Solid line:  a fit of $p(|\dot\theta_{0}|) = S_0/(1+ax^{\varepsilon})$ to the data.   (b): The exponent $\varepsilon$ representing the power-law dependence of the tail of the distribution vs. $R$ for the medium sample (solid circles) and the large sample (open circles).}
\label{fig:prob_dthetadt}                                       
\end{figure}

The instantaneous azimuthal rotation rate $\dot\theta_{0}$ is an important parameter relating to reorientations, and some of its statistical properties are covered in this section.  The probability distribution $p(|\dot\theta_{0}|)$ of the absolute value of $\dot\theta_{0}$ is shown in Fig.~\ref{fig:prob_dthetadt}a for $R=1.1\times 10^{10}$.  The function  $p(|\dot\theta_{0}|) = S_0/(1+ax^{\varepsilon})$ is fit to the data.  The exponent $\varepsilon$, representing the power-law dependence of the tail of $p(|\dot\theta_{0}|)$, is shown for various $R$ in  Fig.~\ref{fig:prob_dthetadt}b. On average it is about $3.6$ and it does not appear to vary much with $R$.  
 
  \begin{figure}                                                
\centerline{\psfig{file=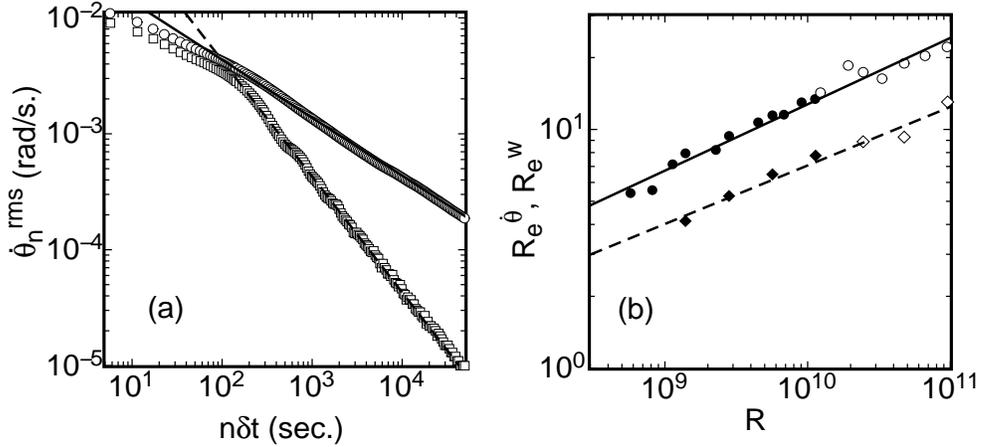,width=5.1in}}
\caption{(a):  Root-mean-square rotation rate $\dot\theta^{rms}_n$ as a function of the  time interval $n \delta t$ for $R=1.1\times10^{10}$ in the medium sample.  Open circles: data for a level sample ($\beta = 0$) with a fit of $\dot\theta^{rms}_n \propto (n \delta t)^{-1/2}$ (solid line).  Open squares: data for a tilted sample (see Sect.~\ref{sec:tilt}) with $\beta = -0.026$ rad with a fit of $\dot\theta^{rms}_n \propto (n \delta t)^{-1}$ (dashed line).  (b):  Reynolds numbers for the rotation rate vs. R.  Circles:  $R_e^{\dot\theta}$ for $\dot\theta^{rms}_n$, with a power-law fit (solid line) yielding the exponent $\chi = 0.278$. Diamonds: $R_e^w$ for the width $w$ of $p(|\dot\theta_{0}|)$, with a power-law fit (dashed line) yielding the exponent 0.245.  Solid symbols: medium sample.  Open symbols: large sample. }
\label{fig:dthetadt_ra}                                       
\end{figure}

The half width at half maximum of $p(|\dot\theta_{0}|)$ is given by $w = a^{-1/\varepsilon}$, but before we can normalize it to get a Reynolds number, we must take into consideration that averaged values of $\dot\theta_{0}$ depend on the experimental sampling interval $\delta t$ because the rotation is dominated by fluctuations. To see this dependence, the root-mean-square rotation rate $\dot\theta^{rms}_n$ was computed for integer multiples of the sampling period $n\delta t$ as $\dot\theta^{rms}_n = \sqrt{\langle[\theta_0(t+n\delta t) - \theta_0(t)]^2\rangle}/(n\delta t)$.  This is shown vs. $n \delta t$ as open circles in Fig.~\ref{fig:dthetadt_ra}a for $R= 1.1 \times 10^{10}$ for a carefully leveled sample ($\beta=0$, see Sect.~\ref{sec:tilt} below).  The function $\dot\theta^{rms}_n = \varphi \cdot  (n \delta t)^{-1/2}$ is shown to fit the data over a large range of $n \delta t$ (solid line). The $(n \delta t)^{-1/2}$ dependence indicates a diffusive process, {\it i.e.} Gaussian distributed fluctuations dominate $\dot\theta^{rms}_n$ in this range of large $n \delta t$. 

We wish to define a Reynolds number $R_e^{\dot\theta} = L^2\dot\theta^{rms}_n/\nu$, but we must choose a $\dot\theta^{rms}_n$ corresponding to a standard time interval, for instance the viscous diffusion time $t_\nu = L^2/\nu$, so that data from different samples can be compared.  This choice yields $\dot\theta^{rms}_{t_\nu} = \varphi \cdot (t_{\nu})^{-1/2}$ and thus $R_e^{\dot\theta} = \varphi L/\sqrt{\nu}$. This Reynolds number is plotted in Fig.~\ref{fig:dthetadt_ra}b vs. $R$ (circles) and a power law $R_e^{\dot\theta} = c R^{\chi}$ is fit to the data (solid line) to obtain $c = 0.0211$ and $\chi = 0.278$.  

We now return to the width $w$ of $p(|\dot\theta_0|)$.  In this case $w$ was calculated only for the experimental sampling interval $\delta t$, but we want to scale it up to $t_{\nu}$ assuming the same $(n\delta t)^{-1/2}$ scaling. Thus we multiply by $\sqrt{\delta t/t_{\nu}}$ and define $R_e^w = L^2w/\nu \times \sqrt{\delta t/t_{\nu}} = w L\sqrt{\delta t/\nu}$. This is also shown in Fig.~\ref{fig:dthetadt_ra}b (diamonds).  This scaling again allows us to compare Reynolds numbers for different experiments with different $L$. Indeed we see that the large and medium samples yield results that follow the same power law. However, $R_e^w$ does not indicate the actual rotation rate over the time scale $t_{\nu}$. The latter is better represented by $\dot\theta^{rms}_n$.  The power law $R_e^w = f R^{\varrho}$ was fit to the data (dashed line) to obtain $f = 0.025$ and $\varrho = 0.245$. We see that $\varrho$ is quite close to $\chi$.

\begin{figure}                                                
\centerline{\psfig{file=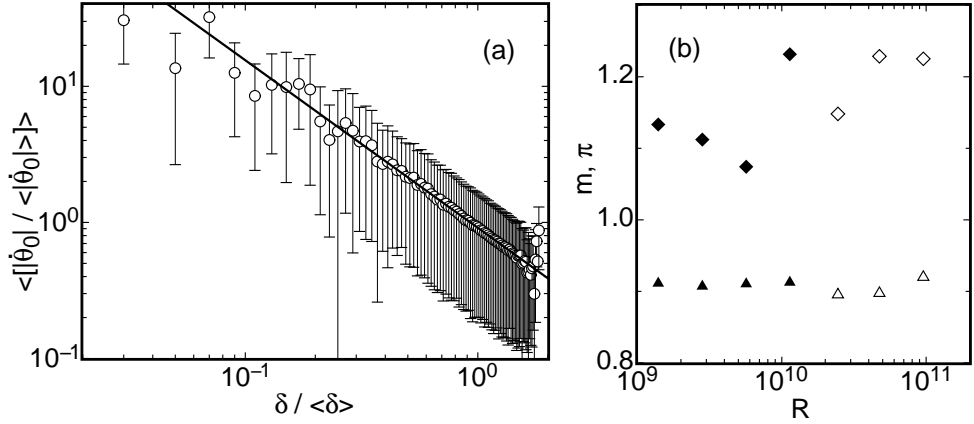,width=5.1in}}
\caption{(a):  The average rotation rate as a function of amplitude for $R = 1.1\times 10^{10}$ in the medium sample with a fit of the power law $\langle|\dot\theta_{0}|/\langle|\dot\theta_{0}|\rangle\rangle = m(\delta/\langle\delta\rangle)^{-\pi}$ to the data (solid line). (b):   The coefficient $m$ (triangles) and exponent $\pi$ (diamonds) vs. $R$ from the power law fit for the medium sample (solid symbols) and the large sample (open symbols).}
\label{fig:dthetadt_amp}                                       
\end{figure}

Next we consider the relationship between the magnitude of the rotation rate $|\dot\theta_{0}|$ and the amplitude $\delta$.   All of the $|\dot\theta_{0}|$ for all $R$ were normalized by the time average $\langle|\dot\theta_{0}|\rangle$ for their $R$ and sorted into bins according to the value of the similarly normalized amplitude $\delta/\langle\delta\rangle$ at the same time step.  The average value of the rotation rate as a function of the amplitude is plotted for each bin as  $\langle|\dot\theta_{0}|/\langle|\dot\theta_{0}|\rangle\rangle$ vs. $\delta/\langle\delta\rangle$ in Fig.~\ref{fig:dthetadt_amp}a for $R=1.1\times 10^{10}$.   The bars represent the sample standard deviation of $|\dot\theta_{0}|$ for each bin, indicating the typical range of $\dot\theta_{0}$.  A fit of the power law$\langle|\dot\theta_{0}|/\langle|\dot\theta_{0}|\rangle\rangle = m(\delta/\langle\delta\rangle)^{-\pi}$ to the data yielded $m = 0.908 \pm 0.009$ and $\pi = 1.16 \pm 0.06$. The same fit was also done separately for several different values of $R$ and the values of $m$ (triangles) and $\pi$ (diamonds) are shown in Fig.~\ref{fig:dthetadt_amp}b. They appear to be independent of $R$.  This analysis shows a nearly inverse relationship between the amplitude and the rotation rate.  While a negative correlation between the two parameters was already evident during cessations, this shows a more general relationship. This is also consistent with a correlation function between $\dot\theta_{0}$ and $\delta$ published before [\cite{BNA05a}], which showed a strong negative correlation between the two parameters.

\section{Tilting the sample}
\label{sec:tilt}

 \begin{figure}
\centerline{\psfig{file=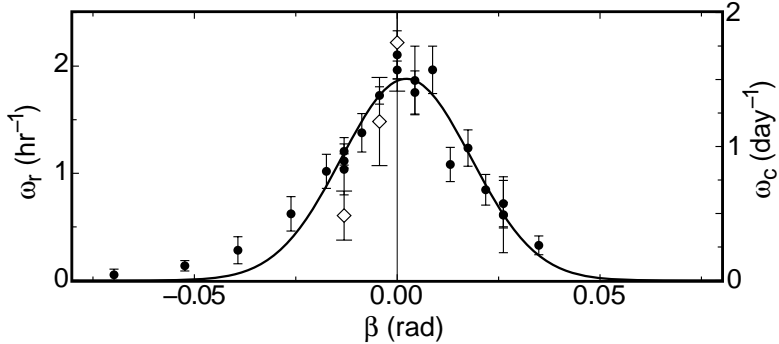,width=4.0in}}
\caption{Solid circles: The frequency of reorientations $\omega_r$ (left ordinate) versus tilt angle $\beta$ form the medium sample at $R=1.1\times 10^{10}$. Solid line:  Gaussian fit to medium sample data.  Open diamonds:  the frequency of cessations $\omega_c$ (right ordinate) vs. $\beta$ in the medium sample at $R=1.1\times 10^{10}$.}
\label{fig:tilt_reor_rate}
\end{figure}

The effects of tilting the sample relative to gravity were investigated before [\cite{ABN05}]. Here we give some additional data relevant to the influence on reorientations. Tilting the sample by an angle $\beta$ breaks the azimuthal symmetry and encourages the fluid in the thermal boundary layers to flow up the bottom plate and down the top plate in the direction $\theta_{\beta}$ of the steepest slope.  Thus, $\theta_{\beta}$ is also the preferred orientation of the LSC in a tilted sample when there are no stronger asymmetries.  First we consider the effect of tilting the sample on the rate of occurrence of reorientations.  Since ${\cal T}$ varies somewhat with $\beta$ [\cite{ABN05,BFA06}], we used the turnover time for $\beta = 0$ to determine $\dot\theta_{min}$ for counting reorientations for all of the tilting experiments.  Figure \ref{fig:tilt_reor_rate} shows the average rate of occurrence of reorientations $\omega_r$ (solid circles) versus tilt angle $\beta$ at $R = 1.1\times 10^{10}$ in the medium sample.  The error bars represent the probable error of the mean.  The analogous data for $R= 9.4\times 10^{10}$ and the large sample was shown by \cite{ABN05}.  For both values of $R$ the reorientations are suppressed significantly even for very small tilt angles.  The Gaussian function $\omega_r = (2\pi\sigma_{\omega}^2)^{-1/2}\exp[-(\beta-\delta\beta)^2/(2\sigma_{\omega}^2) ]$, chosen empirically, was fitted to the data and gave a standard deviation of $\sigma_{\omega} = 0.0160 \pm 0.0006$ rad, indicating how small of a tilt is required to significantly reduce the number of reorientations.  For  $R= 9.4\times 10^{10}$, the standard deviation was  $\sigma_{\omega} = 0.0178 \pm 0.0011$ [\cite{ABN05}], indicating that there is no significant $R$-dependence for the suppression of reorientations by tilt.  The fit of $\omega_r(\beta)$ is symmetric around a center offset of $\delta\beta = 0.0022 \pm 0.0006$ rad, instead of $\delta\beta = 0$ as might have been expected.  We will show elsewhere that this offset is consistent with the influence of the Earth's Coriolis force [\cite{BA06}].

Figure \ref{fig:tilt_reor_rate} also shows the average frequency of cessations $\omega_c$ (open diamonds) versus $\beta$ for three tilt angles where we have more than 10 days of data, all at $R=1.1\times 10^{10}$.  The error bars represent the probable error of the mean.  We also took data for 11 days total at larger tilt angles, with $0.017$ rad $\le |\beta| \le 0.21$ rad, and found no cessations in that period.  Cessations are significantly suppressed in a tilted sample, about as much as reorientations, but because of the scarcity of events we cannot say more on this subject.
 
 The characteristic azimuthal rotation rates were also found to decrease in a tilted sample.  For instance, the instantaneous rotation rate $|\dot\theta_0|$ was reduced to about 50\% of its level-sample value at $\beta = -0.21$ rad.  This is not unexpected given the increase of the amplitude $\delta$ by about 75\% over its level-sample value for $\beta = -0.21$ rad [\cite{ABN05}], and the inverse relationship between the two parameters shown in Fig.~\ref{fig:dthetadt_amp}.   
 
\begin{figure}
\centerline{\psfig{file=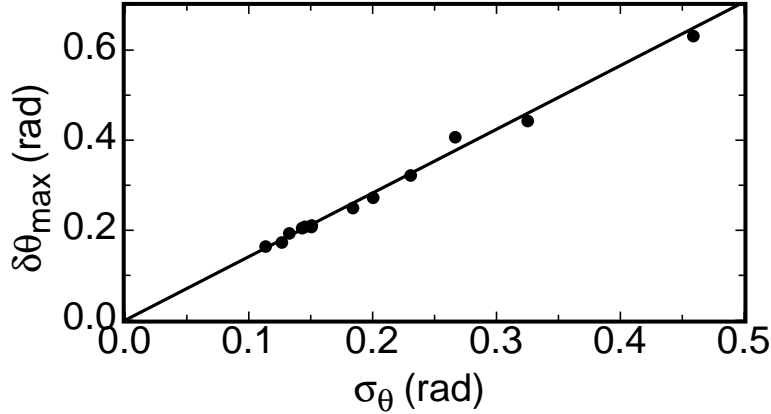,width=4.0in}}
\caption{The saturation value $\delta\theta_{max}$ as a function of  the width $\sigma_{\theta}$ of $p(\theta_0)$, for tilt angles $|\beta| \ge 0.026$ in the medium sample at $R=1.1\times 10^{10}$.  The solid line is the theoretical result $\delta\theta_{max} = \sqrt{2}\sigma_{\theta}$ that should pertain if $\delta\theta_{max}$ is limited by the Gaussian distribution of $p(\theta_0)$. }
\label{fig:tilt_dtheta_ptheta}
\end{figure}

 For more than a minimum tilt angle, the root-mean-square rotation rate $\dot\theta^{rms}_n$ no longer scales as  $(n \delta t)^{-1/2}$ for large $n\delta t$. Rather, for $|\beta| \stackrel{>}{_\sim} 0.026$, it scales as $(n \delta t)^{-1}$ for large $n \delta t$.  This is shown in Fig.~\ref{fig:dthetadt_ra} for $R=1.1\times10^{10}$ and $\beta = -0.026$ (open squares) with a fit of $\dot\theta^{rms}_n = \delta\theta_{max}/(n \delta t)$ to the data, where $\delta\theta_{max}$ is a suggestively named fitting parameter.  The $(n\delta t)^{-1}$ scaling indicates that $\delta\theta^{rms}_n$ saturates at a maximum value $\delta\theta_{max}$ at $n \delta t \approx 300$ s. Presumably this occurs because $\theta_0$ is locked into a small range for large $\beta$, which would suggest that $\delta\theta_{max}$ is related to $p(\theta_0)$.  For tilt angles $|\beta|  \stackrel{>}{_\sim} 0.0087$, $p(\theta_0)$ is fit well by a Gaussian distribution with width $\sigma_{\theta}$ [\cite{ABN05}].  For an ideal Gaussian distribution, the mean-square distance between two points $x$ and $y$ in the distribution is given by 
 $$<(x-y)^2> = \int_{-\infty}^{\infty}dx \int_{-\infty}^{\infty}dy (x-y)^2 (2\pi\sigma_{\theta}^2)^{-1}\exp[-x^2/(2\sigma_{\theta}^2)]\exp[-y^2/(2\sigma_{\theta}^2)]= 2\sigma_{\theta}^2$$  
Thus if the locking of $\theta_0$ into a small range for large $\beta$ is responsible for the saturation of $\delta\theta^{rms}_n$, we should have $\delta\theta_{max} = \sqrt{2}\sigma_{\theta}$.  Figure \ref{fig:tilt_dtheta_ptheta} shows $\delta\theta_{max}$ vs. $\sigma_{\theta}$.  The data point on the upper right is for $\beta = -0.026$ and the point on the lower left is for $\beta = -0.21$.  Also shown in the plot is a solid line for the theoretical prediction $\delta\theta_{max} = \sqrt{2}\sigma_{\theta}$. There is excellent agreement between the two, indicating that the large $n\delta t$ dynamics of $\theta_0$ for $|\beta| \ge 0.026$ are dominated by the tilt locking the orientation into a small range, which overwhelms  the diffusive dynamics that dominate the large $n \delta t$ range for $\beta = 0$.  Even for very small $\beta = -0.0044$ we see deviations from the $(n \delta t)^{-1/2}$ scaling, indicating that these diffusive dynamics only dominate in very symmetric systems.

 It is interesting to consider why reorientations are suppressed by tilting the sample.  If tilting simply added a slight deterministic rotation rate due to the added buoyancy of the boundary layers, then we would expect it to suppress reorientations with $|\dot\theta|$ smaller than the deterministic term.  We can use a highly simplified model to estimate the order of magnitude of this effect. This model is based on one used successfully to estimate the effect of tilting on the Reynolds number [\cite{CRCC04,ABN05}].   Tilting the sample by a small angle $\beta$ relative to gravity results in a buoyancy force per unit area in the thermal boundary layers approximately given by $\rho lg \beta \alpha \Delta T/2$ parallel to the plates, where $l$ is the boundary-layer thickness.  This forcing mainly contributes to enhancing the LSC since it is usually aligned with the slope of the plates [\cite{ABN05}], but when the LSC is not aligned with this slope, some fraction of this forcing will push the LSC into an azimuthal rotation towards the orientation $\theta_{\beta}$. We would expect this fraction to be proportional to $\sin(\theta_{\beta}-\theta_0)$.  This buoyant forcing is opposed by the viscous shear stress from the azimuthal motion across the boundary layer. This opposing force can be approximated by $\rho\nu u_{\theta}/l \approx \rho \nu L\dot\theta_{0}/(2l)$.  Equating these terms, substituting $l = L/(2{\cal N})$ (${\cal N}$ is the Nusselt number), and using the definitions of $R$ and $\sigma$ yields
 
 \begin{equation} 
\dot\theta_{0} = \frac{\beta R\nu\sin(\theta_{\beta}-\theta_0)}{4\sigma {\cal N}^2 L^2}\ .
\label{eq:thetadotofbeta}
\end{equation}
 
 We carried out experiments at $R=1.1\times 10^{10}$ and $\sigma = 4.38$ (${\cal N}=133$, independent of $\beta$ to better than 1\% [\cite{ABN05}]) at various tilt angles. Substituting these values into Eq.~\ref{eq:thetadotofbeta} we obtain $\dot\theta_{0} = 0.40\beta\sin(\theta_{\beta}-\theta_0)$ rad/s.  By expressing $\dot\theta_{0}$ as a function of $\theta_0$ we implicitly assumed that inertia is negligible, i.e. $\ddot\theta_0 = 0$.  This model also ignores possible buoyant forcing in the bulk due to the non-uniform temperature distribution there, and additional viscous stress in the viscous boundary layers as opposed to just the thermal boundary layers which were chosen for the convenience of balancing stresses, as well as any deformation of the LSC due to uneven distribution of these forces. This model is meant to predict a shape for the distribution of $\dot\theta_{0}(\theta_0)$ due to buoyant forces, as well as the order of magnitude of the effect.  Neither of these predictions are expected to change significantly by accounting for the aforementioned imperfections of the model.  
 
 \begin{figure}
\centerline{\psfig{file=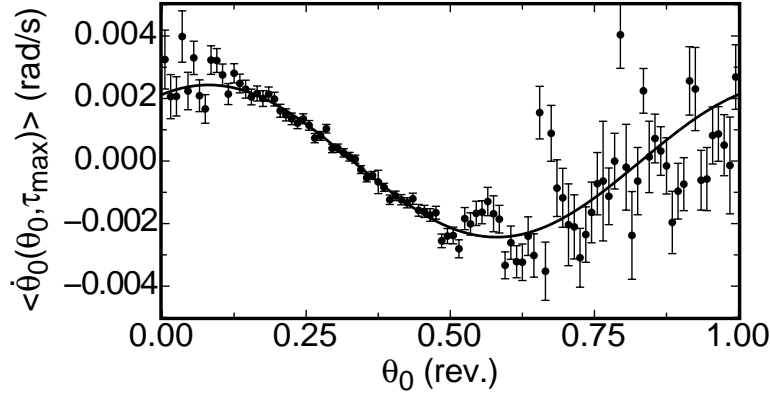,width=4.0in}}
\caption{ The average azimuthal rotation rate $\langle\dot\theta_{0}(\theta_0, \tau_{max})\rangle$ with a delay time $\tau_{max}$ after the LSC has reached the orientation $\theta_0$. Medium sample, $R = 1.1\times  10^{10}$ and $\beta = -0.0044$.  Solid line:  fit of $\langle\dot\theta_{0}(\theta_0,\tau_{max})\rangle = A\sin(\theta_{\beta} - \theta_0)$ to the data.}
\label{fig:dthetadt_theta}
\end{figure}

We can find the magnitude of this effect experimentally by averaging the instantaneous rotation rate $\dot\theta_{0}$ at different orientations $\theta_0$.  Even though the turbulent meanderings of the LSC orientation tend to be much larger than the deterministic part, by averaging over many data points at the same value of $\theta_0$ we are able to resolve the deterministic effects.  The data were binned according to $\theta_0$, and the average azimuthal rotation rate $\langle\dot\theta_{0}(\theta_0)\rangle$ was calculated for each bin.  However, a similar calculation for $\langle\ddot\theta_0(\theta_0)\rangle$ shows that it is proportional to $\sin(\theta_{\beta}-\theta_0)$, and thus that our assumption that $\ddot\theta= 0$ in the above model is not accurate.  If we were to include both the inertial term and a driving term due to turbulence in the above model, then the equation would be equivalent to that of a damped driven pendulum, although we do not know how to represent the driving term.  To better compare the data to the model, we calculated $\dot\theta_{0}(\theta_0,\tau) = [\theta_0(\tau+\delta t/2) - \theta_0(\tau-\delta t/2)]/\delta t$ (where $\tau$ represents a shift of the time axis), and binned the data based on $\theta_0(\tau = 0) = [\theta_0(\delta t/2) + \theta_0(-\delta t/2)]/2$. This yielded a rotation rate as a function of the delay time $\tau$ after $\theta_0$ had been reached.  The above model neglects inertia, so we chose the delay time $\tau = \tau_{max}$ such that $\dot\theta_{0}(\theta_0,\tau_{max})$ is maximized to satisfy $\ddot\theta_0(\tau_{max}) = 0$. Typically $\tau_{max} \approx 100$ s. The results for  $\langle\dot\theta_{0}(\theta_0,\tau_{max})\rangle$  vs. $\theta_0$ for $R = 1.1\times 10^{10}$ and $\beta=-0.0044$ rad are shown in Fig.~\ref{fig:dthetadt_theta}.  The error bars represent the probable error of the mean for each bin. A sinusoidal function $\langle\dot\theta_{0}(\theta_0,\tau_{max})\rangle = A\sin(\theta_{\beta} - \theta_0)$ was fit to the data to find the amplitude $A(\beta)$ of the deterministic rotation rate and preferred orientation $\theta_{\beta}(\beta)$.  It should be noted that most of the time $\theta_0$ is near the preferred orientation (where $\langle\dot\theta_{0}(\theta_0,\tau_{max})\rangle = 0$ and $\nabla\langle\dot\theta_{0}(\theta_0,\tau_{max})\rangle < 0$), and thus the error bars are much smaller near this orientation in Fig.~\ref{fig:dthetadt_theta}.  For $|\beta| \stackrel{>}{_\sim} 0.027$, not all orientations are sampled and the fit to $\langle\dot\theta_{0}(\theta_0,\tau_{max})\rangle$ is essentially a linear fit near $\theta_{\beta}$.

 \begin{figure}
\centerline{\psfig{file=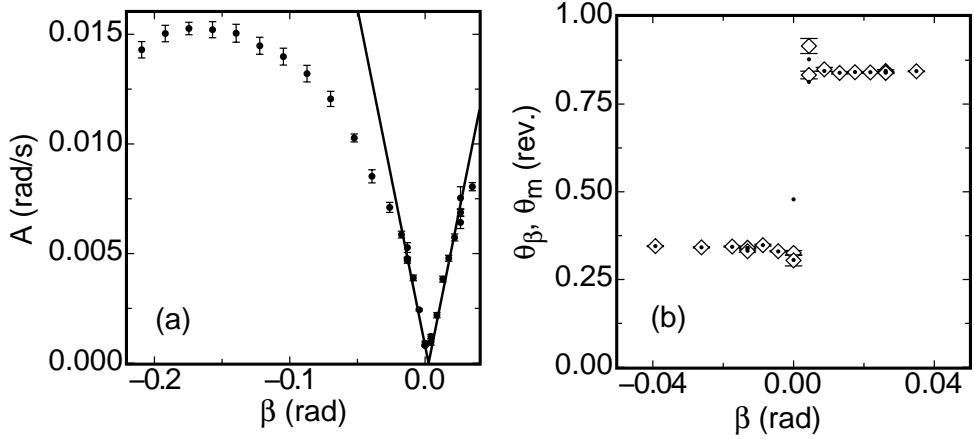,width=5.1in}}
\caption{(a):  The coefficient of $A(\beta)$ from $\langle\dot\theta_{0}(\theta_0,\tau_{max})\rangle = A(\beta)\sin(\theta_{\beta} - \theta_0)$.  Solid line:  a fit of $A(\beta) = A'|\beta-\beta_0|$ to the data for $|\beta | \le 0.027$. (b):  The preferred orientation $\theta_{\beta}$ obtained from the fit (open diamonds) and $\theta_m$ from the peak of $p(\theta_0)$ (small circles). Data is for $R=1.1\times10^{10}$ in the medium sample.}
\label{fig:dthetadt_tilt}
\end{figure}

The fit parameter $A(\beta)$ is shown in Fig.~\ref{fig:dthetadt_tilt}a and $\theta_{\beta}(\beta)$ (open diamonds) is shown in Fig.~\ref{fig:dthetadt_tilt}b, for several tilt angles $\beta$ at $R=1.1\times 10^{10}$ in the medium sample.  Also plotted in Fig. \ref{fig:dthetadt_tilt}b is the preferred orientation $\theta_m$ (small circles) obtained from the peak of $p(\theta_0)$.  As we would expect, both methods of finding the preferred orientation agree with each other, so: $\theta_m  = \theta_{\beta}$, at least for tilt angles $|\beta| > 0.01$ rad.  Figure \ref{fig:dthetadt_tilt}a shows a linear fit of $A(\beta) = A'|\beta-\beta_0|$ for $|\beta| < 0.026$, which yields $A' = 0.307 \pm 0.002$ rad/s and $\beta_0 = 0.00258 \pm 0.00008$ rad.  These plots show that $\langle\dot\theta_{0}(\theta_0,\tau_{max})\rangle \propto \beta\sin(\theta_{\beta} - \theta_0)$ as was predicted for $\beta < 0.026$, although the proportionality breaks down for larger $\beta$.  The calculated coefficient from the model of $0.40$ rad/s is slightly larger than the fitted value of $A'$, which we consider to be good agreement.  This shows that the addded buoyancy in the tilted sample is directly responsible for the measured deterministic rotation rate for $|\beta| \le 0.026$.  The fact that $\beta_0 \ne 0$ is again due to the Earth's Coriolis force [\cite{BA06}].

Finally we consider what effect the $\beta$-dependence of the deterministic rotation rate has on reorientations.  The minimum rotation rate for counting reorientations for these data is $\dot\theta_{min} = 0.0254$ rad/s, which is much larger than the deterministic forcing $A =  0.0047$ rad/s found for $\beta = -0.013$, and it is even larger than the largest measured value for the deterministic forcing $A = 0.015$ rad/s.  Thus we conclude that the direct forcing from tilting the sample that is considered in the above model is too small to account for the 40\% reduction in reorientations at $\beta = -0.013$ relative to the level sample and the complete suppression of reorientations within our resolution at larger values of $|\beta|$.  There must be some other mechanism for the suppression of reorientations that we have not yet identified.

\section{Comparison with contemporary experiments}
\label{sec:Xi}

In a set of contemporary experiments, \cite{XZX06} report measurements of the orientation of the LSC for a RBC sample with $\Gamma = 1$, $\sigma \approx 5$, and $10^9 < R < 10^{10}$.  To study short-term dynamics, they used particle image velocimetry to visualize the horizontal fluid velocities near the top plate. From these measurements they calculated a spatially-averaged velocity and orientation for the LSC.  For long-term measurements, they used a bead attached to a ``fishing line" near the bottom plate to determine the orientation of the LSC.

Many of the results reported by \cite{XZX06} are similar to ours and provide an excellent complement to our measurements, as they use velocity measurements and we use temperature measurements to quantify aspects of the LSC.  Some of the results deserve special comment in relation to our work.

\cite{XZX06} report an oscillation of the LSC orientation around a preferred orientation that they measured near the top and bottom plates.  This oscillation was measured before by \cite{FA04} on the basis of plume motion across the top and bottom plates. We find it as well, but only in the upper and lower rows of thermistors at heights $3L/4$ and $L/4$. As seen  by \cite{FA04}, the oscillations in the top and bottom rows are out of phase with each other. This can be seen to some extent in Fig.~\ref{fig:reorientation_examples} (we will report on it in detail elsewhere [\cite{BFA06}]).   We note that other aspects of the LSC azimuthal dynamics can depend quantitatively on the height at which they are measured, but so far this is the only process  known to have such a qualitative height dependence.

 \begin{figure}
\centerline{\psfig{file=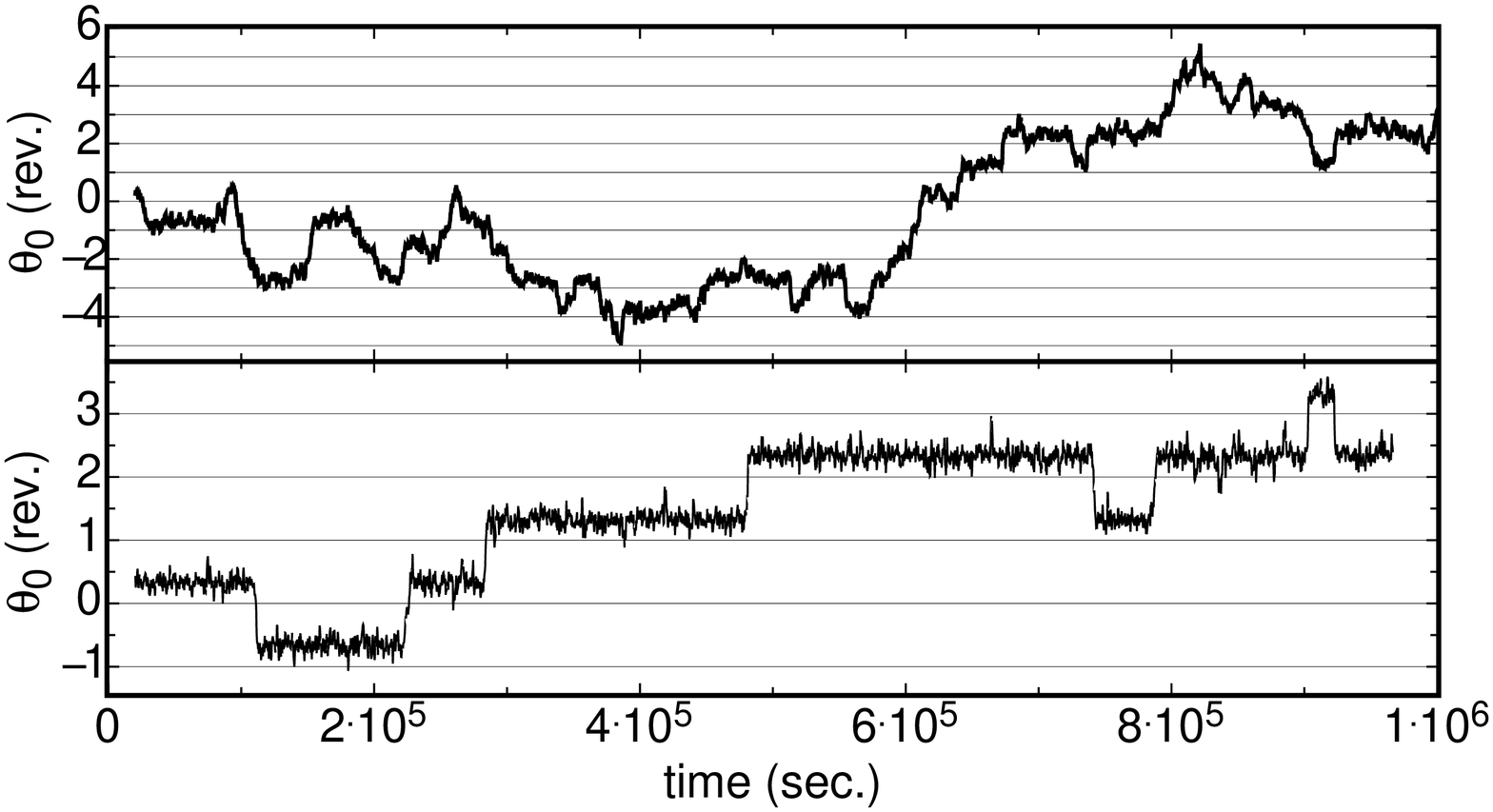,width=4.0in}}
\caption{The orientation $\theta_0(t)$ of the LSC over 11.6 days for $R=1.1\times 10^{10}$ in the medium sample.(a):  $\beta = 0$ and (b): $\beta = -0.0044$ rad.}
\label{fig:theta_t_tilt}
\end{figure}

\cite{XZX06} report that the LSC orientation remains locked near a preferred angle for the duration of an experimental run for a large majority of such runs.
Curiously, this preferred orientation changes with each experimental run in a non-reproducible manner and thus  cannot be due entirely to some geometrical asymmetry of their apparatus. It suggests a more complicated symmetry-breaking process.  For comparison to our experiments we show two time series of $\theta_0(t)$ over 11.6 days at $R=1.1\times 10^{11}$ in Fig.~\ref{fig:theta_t_tilt}.  The upper plot is for $\beta=0$ and the lower one is for $\beta=-0.0044$ rad.  Consistent with the results shown in Fig.~\ref{fig:prob_theta}, the  $\beta=0$ data show a weak preference for some angle, presumably due to asymmetries such as the Coriolis force or very slight deformations of the side walls [\cite{BA06}]; but the preferred orientation is not nearly as severe as the one observed by \cite{XZX06}.  The plot for $\beta = -0.0044$ rad looks qualitatively similar to the time series reported by \cite{XZX06}, although the frequency of rotations through an entire revolution is much lower and $p(\theta_0)$ is still wider in our case.  Since in our case the asymmetry is due to tilting the sample, we should not expect these statistical quantities to correspond exactly to those of \cite{XZX06}.

Considering the differences between the two experiments, the cause for this preferred orientation may be the sapphire top plate in their experiment, with about 1/10 the conductivity of our copper plates.  Direct numerical simulations (DNS) of the Boussinesq equations by \cite{Ve04} and experiments by \cite{BNFA05} show that low-conductivity top and/or bottom plates reduce the heat transport across the sample. The DNS suggest that the low plate conductivity reduces the plume emission from the plates.  Following Verzicco, we suggest that it is possible that the  LSC put a significant thermal imprint on the sapphire plate.  This thermal imprint would diffuse more quickly in a copper plate due to its higher conductivity. The thermal imprint in the sapphire plate presumably discourages cold plume emission, thus introducing a preferred orientation of the circulation corresponding to the initial orientation chosen by the LSC.  If the thermal imprint remains longer than the duration of cessations and full azimuthal revolutions of the LSC, then the preferred orientation can remain so for the duration of the experimental run.  This would also explain why the preferred orientation changes between runs, assuming the flow is stopped for some significant length of time between them.

 \begin{figure}
\centerline{\psfig{file=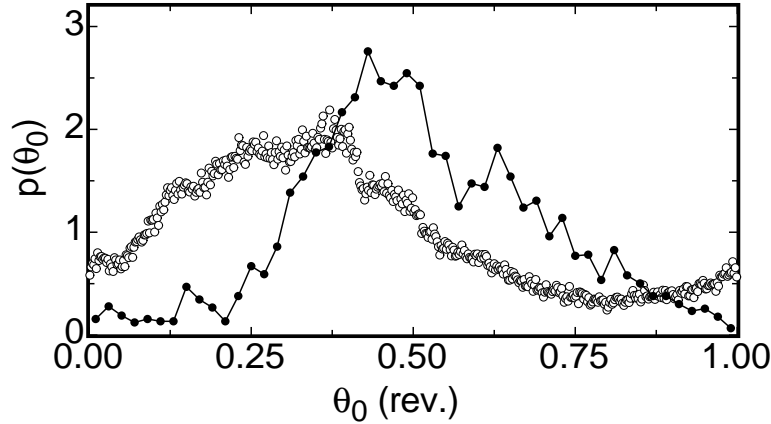,width=4.0in}}
\caption{The probability distribution $p(\theta_0)$ for $R = 1.1\times 10^{10}$ in the medium sample for copper plates (open circles) and aluminum plates (solid circles). }
\label{fig:prob_theta_plate}
\end{figure}

To test this model, we examined data taken with aluminum top and bottom plates [\cite{BNFA05}] in place of the usual copper ones.  Aluminum has a thermal conductivity about 1/2 that of copper, which is still a factor of about four or five larger than that of sapphire, but since the aluminum plates slightly suppressed the overall heat transport relative to the copper-plate system, we hoped to see some influence also on the LSC orientation.  In this case, the eight side-wall thermistors were glued to the outside of the side wall at a height $L/2$, so they were $0.31$ cm away from the fluid. The thermal response time through the side wall is expected to be $d^2/\kappa_{sw} = {\cal O}(10)$ s.  The locking of the orientation is compared in Fig.~\ref{fig:prob_theta_plate} by $p(\theta_0)$ for copper plates (open circles) and aluminum plates (solid circles).  The distribution is more strongly peaked with aluminum plates and does not sample all angles as well as with copper plates , indicating that the orientation tends to be more strongly locked in place with aluminum plates.  This supports our model about the thermal imprint suppressing orientation changes of the LSC, but a pair of experiments comparing copper and sapphire plates with a much larger ratio of conductivities would be more convincing.  

\cite{XZX06} report finding a double-cessation event, in which there are two successive cessations, and the second cessation returns the LSC to its original orientation.  While we have not identified these events in our system, they are not necessarily inconsistent with our measurements.  Since we count a cessation as occurring for the duration that the amplitude $\delta$ remains below some threshold $\delta_h$, if a double cessation occurred in our system, we would most likely count it as a single cessation.  Our measurements of $\delta$ have a relative error of about 12\% when $\delta$ is near its average value, making it difficult for us to resolve double cessations.  However, it is also quite possible that the double cessations are uniquely due to the suppression of plumes due to the thermal imprint on the sapphire plate, which could explain why the LSC is forced back into its original orientation.

The average frequency of cessations (also counting double cessations) for the data reported by \cite{XZX06} is $\omega_c = 1.7$ day$^{-1}$ with a probable error of the mean of $0.3$ day$^{-1}$.  This is consistent with our value of $1.5 \pm 0.1$ day$^{-1}$.  Considering the different types and locations of the measurements of the two experiments, and the sensitivity of cessation frequency to minor asymmetries as well as to how they are counted, it is somewhat fortuitous that the frequency of events in both experiments is in such good agreement.  \cite{XZX06} report an increasing frequency of cessations with increasing $R$, in contrast to our uniform frequency of cessations.  This could be due to the criteria used for defining cessations:  we used a minimum amplitude $\delta_l$ that changes with $R$, while \cite{XZX06} use the velocity time series to find cessations.  While we expect the velocity and temperature amplitudes to have similar behavior, we do not know if they should behave exactly the same during cessations, and thus we do not know if our methods of counting cessations are fully equivalent.

The results from the analysis of crossings by \cite{XZX06} (they referred to the events as ``reversals") are qualitatively similar to ours. In particular, the shape of the probability distribution of the time intervals $\tau_1$ between crossings has the same shape (see our Fig.~\ref{fig:crossings_rev}).  Notably, the time interval at the crossover between the power-law dependence and the exponential dependence is about $10{\cal T}$ in both cases.  The characteristic decay time of the exponential region was significantly larger for \cite{XZX06} ($54{\cal T}$) than for us ($32{\cal T}$). However, since this represents a typical long time interval between crossings, and the azimuthal motion was suppressed in their case, this difference is not surprising.

\section{Summary and conclusions}
\label{sec:summary}

We presented a broad range of measurements of the orientation of the LSC, including rotations and cessations.  These events have not been well-studied experimentally or theoretically in the past, and we have very little physical understanding of how these phenomena occur.  One important conclusion we can make is that when the LSC slows to a stop during a cessation, it loses information of its previous orientations and restarts at a random new orientation. We also found that both cessations and reorientations have a Poisson distribution in time, indicating the independence of successive events. 
We measured the rate of change of the amplitude $\dot\delta$ during cessations. Its value can be compared with future dynamical theories of cessations.

\cite{SBN02} found that the time interval $\tau_1$ between their successive events had a power-law probability-distribution with an exponent of minus one when $\tau_1$ was small, and was cut off exponentially at larger $\tau_1$. They, and in more detail \cite{SBN04}, interpreted the power law as indicative of  self-organized criticality (SOC) and attributed the exponential cut-off to a finite-size effect. In another paper \cite{HYBNS05} proposed an analogy between the statistics of wind reversals on the one hand and that of fluctuations of the magnetization of the two-dimensional Ising model on the other. Again this analogy rested heavily upon the existence of a power-law probability-distribution for $\tau_1$. Our results for the statistics of the rotations and cessations that make up our reorientations are inconsistent with this interpretation and reveal a purely Poissonian probability distribution for our data which goes to a constant at small $\tau_1$ and drops off exponentially at  large $\tau_1$; but when we include the relatively small-amplitude and high-frequency jitter of $\theta_0(t)$ in the analysis by considering ``crossings", then we reproduce the statistics observed by \cite{SBN02}. Thus we conclude that the SOC and the Ising-model analogy discussed by them and by \cite{HYBNS05} can perhaps apply to the jitter of our measurements of $\theta_0(t)$ but does not pertain to the reorientations ({\it i.e.} rotations and cessations) observed by us. We also have no reason to invoke a finite-size effect to explain a large-$\tau_1$ cutoff because the Poissonian statistics of our reorientations naturally yields exponential behavior at large $\tau_1$.

Both reorientations and cessations are found to be strongly suppressed in a tilted sample. The measurements could not be accounted for simply by the buoyancy of the boundary layer creating a preferred orientation of the LSC. This leaves an open problem for future work to answer.  In addition it is unknown whether this reduction in the frequency of events applies to other types of asymmetries, such as a strong Coriolis force or more complicated geometries that are common in geophysical systems.

We presented data  showing a statistical relationship between the rotation rate $|\dot\theta_{0}|$ and the amplitude $\delta$ of the LSC.  This is interesting because of a phenomenon found in nature: reversals in the orientation of the Earth's magnetic field presumably as a result of convection reversal in the outer core are also known to be accompanied by a decrease in the amplitude of the resulting magnetic field [\cite{GCHR99}].   Previously \cite{BNA05a} reported a correlation function of $\delta$ and $|\dot\theta_{0}|$ in which $\delta$ tends to lead $|\dot\theta_{0}|$ by about 6\% of the turnover time.  If this is extrapolated to the 400 year turnover time of the Earth's outer core, measurements of the Earth's magnetic-field amplitude could give several years' notice of a magnetic-field reversal.  Convection in the Earth's core is in many ways different from our ideal convection experiment, so this extrapolation is highly speculative; but it is a subject worthy of future study if it could lead to a forecasting of events which would have a significant impact on human life.

\section{Acknowledgments}

We are grateful to Alexei Nikolaenko for his contribution to the experimental work.  We are also grateful to Detlef Lohse, Penger Tong, and Ke-Qing Xia for stimulating discussions.  This work was supported by the US Department of Energy through Grant  DE-FG02-03ER46080.

\section{Appendix:  The velocity and temperature distributions in the plume layer}

The bimodal vertical velocity distribution near the side wall reported by \cite{NSSD01} led them to suggest that the LSC orientation switched between two opposite directions aligned with their sensors.  We suggest that a bimodal local velocity distribution does not imply a bimodal probability distribution $p(\theta_0)$, and that the orientation could have been varying erratically through any azimuthal angle in their experiment.  

Recent particle velocimetry measurements in a cylindrical RBC sample of $\Gamma = 1$ and $R = 7.0\times 10^9$ provide some useful information [\cite{SXT05}].  For the vertical velocity near the side wall in the plane aligned with the preferred orientation of the LSC (caused by a slight tilt), no reversals of the velocity direction were found.  This agrees with our results for a tilted sample discussed in Sect.~\ref{sec:tilt}. In the plane orthogonal to this, many random velocity reversals were found, and they happened at the same time on opposite sides of the sample, so that there was always one side with up-flow and one side with down-flow  Further, the magnitude of the velocity averaged over short time intervals (10 min, without reversals) was about the same in both planes.  This orthogonal-plane result is qualitatively similar to that of  \cite{NSSD01}, although the latter did not have a tilted sample and did not know the preferred orientation of the LSC in their system.  Since \cite{SXT05} found no velocity reversals in the plane of the LSC, it follows that the observed reversals in the orthogonal plane were not likely to be global reversals of the LSC.  Since the magnitude of the velocity was about the same in both planes, this suggest that the vertical velocity distribution near the side wall was nearly a step function with velocity $+v$ at angles within $\pm \pi/2$ of $\theta_0$ and $-v$ at all other angles, although realistically the velocity must smoothy transition from $+v$ to $-v$ over some small angular range. Since \cite{NSSD01} only observed local velocity reversals, these could have mostly been small orientation changes in which the $+/-v$ interface moved past the sensors.  If the sample of \cite{NSSD01} was sufficiently symmetric, it is still likely that some of the measured events were rotations or cessations.

 \begin{figure}
\centerline{\psfig{file=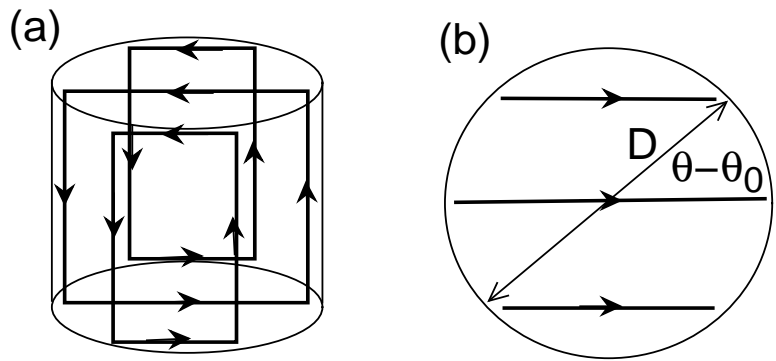,width=4.0in}}
\caption{(a):  A model of the velocities in the outer shell of the LSC. (b): top view of the bottom plate, showing the length $L$ of streamlines crossing the plate. Each streamline is referred to by the azimuthal angle $\theta$ where it intersects with the side-wall boundary-layer. }
\label{fig:vel_model}
\end{figure}

The question remains whether this step-function velocity-distribution can be reconciled with our measurement of the shape of the temperature distribution around the side wall, which we found to be given by $T(\theta) =  T_0 + \delta T$ with $\delta T =  \delta \cos(\theta - \theta_0)$ (see Fig.~\ref{fig:temp_theta}).  Here we present a model of the thin shell between the boundary layer and the bulk, also called the plume layer.   We assume that the average flow in this shell is parallel to the nearest wall of the container, and  that the streamlines make complete loops and are parallel to the plane of the LSC as drawn in Fig.~\ref{fig:vel_model}.  This time average should be thought of as along enough to identify the LSC but short compared to the time between reorientations.   Our calculation is based on the steady-state heat-transport equation

$$(\vec u \cdot \vec\nabla)T = \kappa \nabla^2 T\ .$$

\noindent We approximate the solution for the horizontal plume layers near the top and bottom plate of the sample, assuming that there the  heat transfer to the plume layers  is dominated by contact with the thermal boundary layers.  The transport term is given by $(\vec u \cdot \vec\nabla)T = u dT/dx \approx 2u \delta T/\delta x = 2u(\theta) \delta T(\theta)/|D\cos(\theta-\theta_0)|$ where we approximated the temperature gradient in the plane of the LSC by the horizontal temperature difference $2\delta T(\theta)$ over the path length $\delta x = D\cos(\theta-\theta_0)$.  The heat flux to (from) the bottom (top) plume layer from the thermal boundary layer should not depend on $\theta$, so $\nabla^2 T  \approx +(-) constant$.  Putting these terms together yields the desired relationship between the horizontal velocity and temperature:

$$ u(\theta) \delta T(\theta) \propto +(-)\cos(\theta-\theta_0)$$

\noindent We found a horizontal temperature difference of the form $\delta T(\theta) = \delta\cos(\theta-\theta_0)$, which on the basis of our model implies that $u(\theta)$ has a uniform magnitude but opposite direction near the top and bottom boundary layers.  Assuming that viscous drag and other mechanisms do not significantly affect the plume-layer velocity, following the streamlines in Fig.\ref{fig:vel_model}a implies that the vertical velocities along the side wall must have the same magnitude, resulting in the previously mentioned step-function velocity-distribution.  This model ignores many aspects of the LSC, and thus cannot be a complete description of it.  The model is meant to provide an understanding of why our measured temperature distribution around the side wall has a different shape than the velocity distribution suggested by the data of \cite{SXT05}.  Additionally it justifies our suggestion of the step-function velocity-distribution in discussing the data of \cite{NSSD01}.  Without the qualitative features of the model described here it might be difficult to reconcile the results of the three experiments.

\end{document}